\documentclass[double]{article}
\usepackage{times}
\usepackage{algorithm}
\usepackage{algorithmic}
\usepackage{wrapfig}
\usepackage{verbatim}
\usepackage{amsmath}
\usepackage{graphicx}
\usepackage{subcaption}
\usepackage[active]{srcltx}
\usepackage{color}
\usepackage{lineno}
\usepackage{dirtytalk}
\usepackage{amsthm}
\usepackage{authblk}
\usepackage{listings}
\usepackage{tikz}
\usetikzlibrary{shapes.geometric, arrows}

\usepackage{epsfig,graphicx,amsfonts}
\usepackage{amsmath,amsthm,amssymb}
\usepackage{xcolor}

\usepackage{adjustbox}
\usepackage{multirow}
\usepackage{setspace}
\usepackage{hyperref}
\usepackage{comment}

\long\def\comment #1\commentend{}

\begin{document}
%\documentstyle[amsfonts]{article}

% Following creates the title page
\title{\Large Individual Variation Affects Outbreak Magnitude and Predictability in an Extended Multi-Pathogen SIR Model of Pigeons Vising Dairy Farms}

\author{Teddy Lazebnik$^{1*}$, Orr Spiegel$^{2}$\\
\(^1\) Department of Cancer Biology, Cancer Institute, University College London, London, UK\\
\(^2\) School of Zoology, Faculty of Life Sciences, Tel Aviv University, Tel Aviv, Israel\\

\(*\) Corresponding author: lazebnik.teddy@gmail.com 

}

\date{ }

\maketitle 

\begin{abstract}
Zoonotic disease transmission between animals and humans is a growing risk and the agricultural context acts as a likely point of transition, with individual heterogeneity acting as an important contributor. Livestock often occurs at high local densities, facilitating spread within sites (e.g. among cows in a dairy farm), while wildlife is often more mobile, potentially connecting spatially isolated sites. Thus, understanding the dynamics of disease spread in the wildlife-livestock interface is crucial for mitigating these risks of transmission. Specifically, the interactions between pigeons and in-door cows at dairy farms can lead to significant disease transmission and economic losses for farmers; putting livestock, adjacent human populations, and other wildlife species at risk. In this paper, we propose a novel spatio-temporal multi-pathogen model with continuous spatial movement. The model expands on the Susceptible-Exposed-Infected-Recovered-Dead (SEIRD) framework and accounts for both within-species and cross-species transmission of pathogens, as well as the exploration-exploitation movement dynamics of pigeons, which play a critical role in the spread of infection agents. In addition to model formulation, we also implement it as an agent-based simulation approach and use empirical field data to investigate different biologically realistic scenarios, evaluating the effect of various parameters on the epidemic spread. Namely, in agreement with theoretical expectations, the model predicts that the heterogeneity of the pigeons' movement dynamics can drastically affect both the magnitude and stability of outbreaks. In addition, joint infection by multiple pathogens can have an interactive effect unobservable in single-pathogen SIR models, reflecting a non-intuitive inhibition of the outbreak. Our findings highlight the impact of heterogeneity in host behavior on their pathogens and allow realistic predictions of outbreak dynamics in the multi-pathogen wildlife-livestock interface with consequences to zoonotic diseases in various systems.  \\ \\

\end{abstract}

\maketitle \thispagestyle{empty}

% Begin using page numbers and a header
\pagestyle{myheadings} \markboth{Draft:  \today}{Draft:  \today}
\setcounter{page}{1}% reset page number to 1

\section{Introduction}
\label{sec:introduction}
Disease ecology and animal movement ecology are inherently linked as animal movement can both determine pathogens’ spread and be influenced by their load \cite{intro_1,intro_2,intro_0}. These interfaces have direct and indirect links to the pandemic spread across species in general, and between animals and humans, in particular \cite{intro_3}. As humanity spreads, more places are becoming urban which inherently changes the environment and the biodiversity of the area \cite{intro_4}. Specifically, livestock and synanthropic wildlife that live next to humans (e.g., cows and pigeons, respectively) have a complex relationship among themselves and with humans. For example, pigeons commonly occupy urban and agricultural sites worldwide and are a known vector of various human, poultry, and livestock-relevant pathogens \cite{intro_5,intro_5b}. Nonetheless, our understanding of multi-species epidemic spread dynamics, in general, and in the context of mixed urban and agricultural sites with individuals moving between them is limited. Specifically, current models are not designed to capture the unique multi-pathogen infection which can take place in parallel on the individual level and is likely the common case in most systems. Moreover, wild animals usually follow an exploration-exploitation movement pattern while captive animals are constrained to a small spatial area (e.g. a farm). The influence of this unsymmetrical movement dynamics on an epidemic spread is not fully explored.

The investigation of interacting species has gained significant popularity, leading to the continuous unveiling of the biological dynamics that surround us, while also serving as a fundamental basis for various technological advancements \cite{tech_1,tech_2,tech_3,tech_4}. Particularly, there has been a notable focus on studying epidemiology to comprehend the transmission of infectious diseases. The ultimate aim is to devise effective pandemic intervention strategies and, ideally, eliminate these diseases altogether, or more proximately, prevent them on a local scale \cite{pip_1,pip_2,pip_3,pip_4,pip_5,building_sir_ariel}. In this regard, mathematical models and computer simulations have emerged as potent tools for comprehending the biological and ecological dynamics that underlie the observed patterns associated with the spread of pandemics \cite{powerful_tool_1,powerful_tool_2,powerful_tool_3,powerful_tool_4, alexi2022trade,spatio_temporal_airborne_andemic_ariel}. Multiple attempts have been proposed to model the spread of epidemics in populations \cite{stat_model_1,different_approach_from_sir,different_approach_from_sir_2,models_good_1,models_good_2,models_good_3}. In particular, original Susceptible-Infected-Recovered (SIR) based models have been improved and extended to models that offer more realistic spatial, social, biological, and other dynamics compared to the original SIR. These models are now widely used due to their balancing between representation simplicity and prediction accuracy \cite{new_review_1,new_review_2,review_3}. Initially, extended SIR models with a single species and a single pathogen were proposed and investigated \cite{multi_populations_1,multi_populations_2,multi_populations_3}. For instance, Sah and colleagues (2017) \cite{or_sir_example} used SIR models to show the effect of social structure and network modularity on the outbreak dynamic, demonstrating with empirical social network data from 43 species how group living can actually slow down simulated outbreaks in some conditions. \cite{new_review_2} proposed a highly detailed, stochastic, and spatio-temporal extended SIR model for disease progression in animal-to-animal contact. The authors implemented their model as a computer simulation, allowing users to explore a wide range of possible pandemic intervention policies. 

Later studies widened these models by taking into consideration multi-strain and even multi-pathogen pandemics \cite{bio_example_1,sir_mutation_4,teddy_chaos}. For instance, \cite{dang} proposed a multi-strain model that links immunological and epidemiological dynamics across scales where within the host scale, the strains eliminate each other with the strain having the largest immunological reproduction number persisting and on the population scale, the authors adopted an extended SIS (Susceptible-Infected-Susceptible) model. \cite{multi_strain_3} focused on a two-strain SEIR (E - exposed) model with dynamics infection rates. The authors show that this model is able to capture the dynamics of an emerging second strain during a pandemic. They demonstrated this capability using data from the 2020 COVID-19 pandemic. To this end, because pathogens co-occurring pathogens can influence each other through modification of host behavior, physiology, and survival, researchers are interested in extending these models for multiple species and the interactions between them \cite{beb_1,pp_equlibrium_1,hard_msms_4}. \cite{base_paper} proposed a multi-strain multi-species extended SIR model where the authors combined the pray-predator model \cite{lv_example_1,lv_example_2} with an extended version of the multi-strain SIR model proposed by \cite{teddy_multi_strain}. The model allows them to evaluate the extinction of species due to natural pandemics, using only macro data (i.e., average over the population, and not animal-level data) about the animal's prey-predator dynamics and cross-infection for the Avian influenza pathogen and its strains. Adding realism to the model requires focusing on a multi-location scenario, in addition to the multi-pathogen consideration. 

\begin{figure}[!ht]
    \centering
    \begin{subfigure}{.49\textwidth}
        \includegraphics[width=0.99\textwidth]{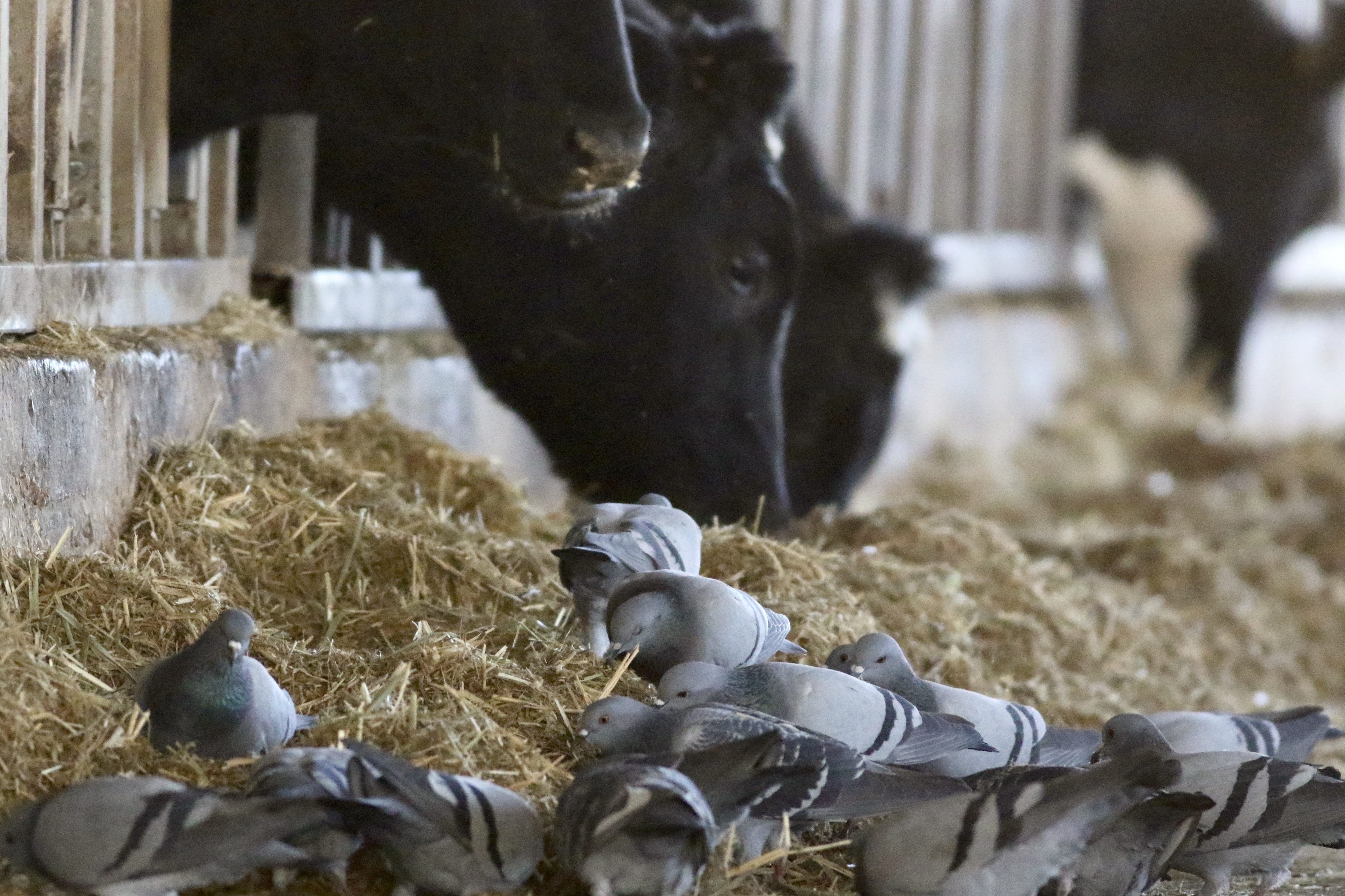}
        \caption{Pigeons-cows proximity in a cowshed.}
        \label{fig:atmo_1}
    \end{subfigure}
    \begin{subfigure}{.49\textwidth}
        \includegraphics[width=0.99\textwidth]{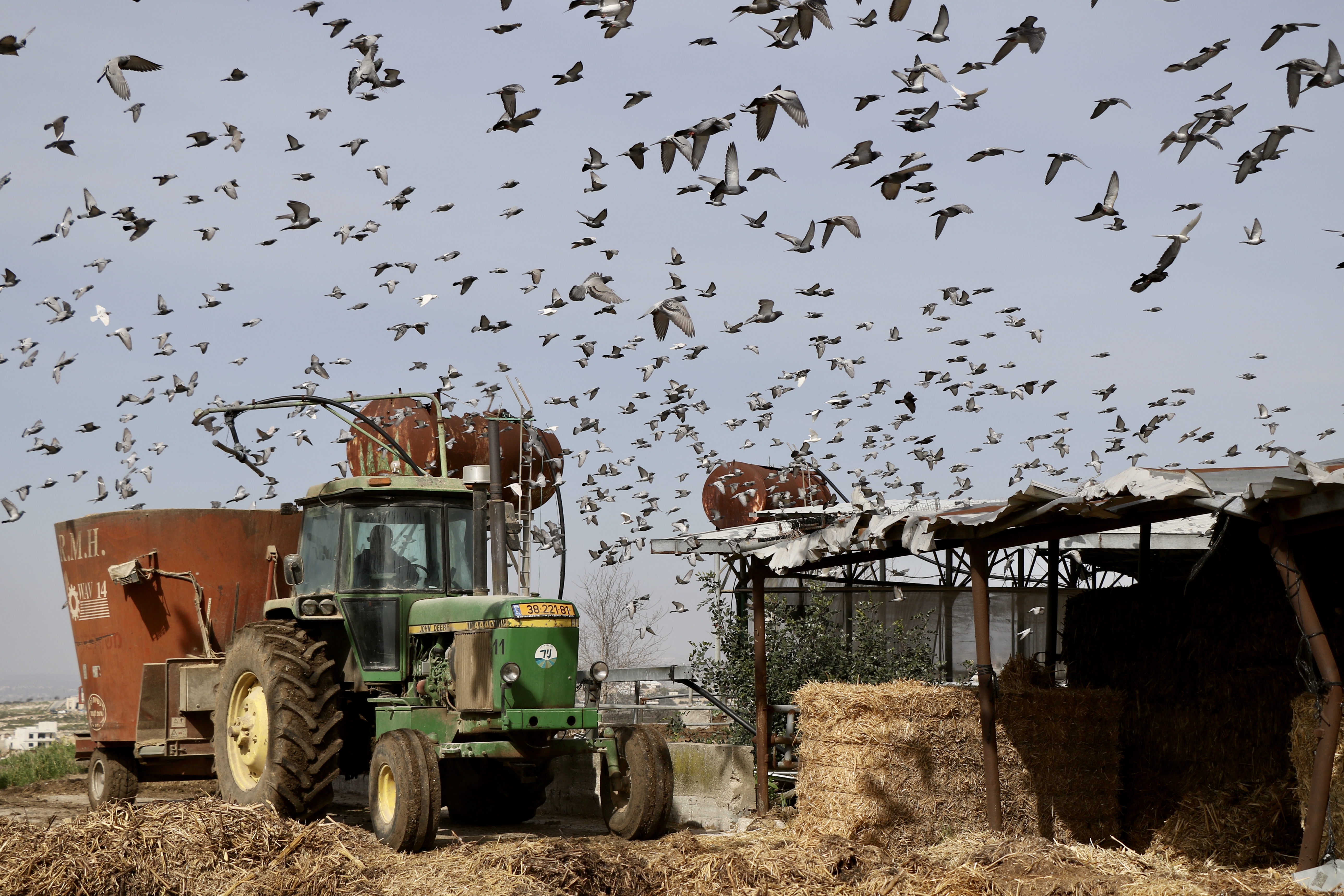}
        \caption{Pigeons fly towards a cowshed.}
        \label{fig:cowsheds_mi}
    \end{subfigure}
    
    \caption{Two images of pigeons in a cowshed located in Israel,  taken on the site of the empirical data collection of  \cite{orr_base_paper}. The images show the high proximity of pigeons to cows and their food (facilitating transmission), and the high density and mobility of pigeons (underscoring their between-site transmission potential). Photos by Tovale Solomon. }
    \label{fig:atmo}
\end{figure}

Pigeons visiting dairy farm \cite{pigeons_foo_1,pigeons_foo_2}. are a typical and common example of such multi-site, multi-pathogen, multi-hosts scenarios, with mixed urban and agricultural sites. One cause of interest is their interactions with livestock, in general, and cows in cowsheds in particular \cite{intro_6,intro_7}. Importantly, like many bird species, pigeons are highly social animals and they tend to aggregate in large flocks which previous studies pointed out as a key factor in the spillover and worldwide spread of Avian Influenza \cite{pigeons_human_2}. Moreover, dairy and poultry farming are particularly notorious for attracting a variety of wildlife from peripheral habitats, due to their resource availability including food, shelter, and possible security from natural predators who avoid human habitats \cite{pigeons_human_1}. In this context, pigeons are able to cover large areas flying several kilometers on a daily basis \cite{intro_8}, effectively operating as an infection vector across otherwise spatially separated sites. Generally speaking, pigeons travel between an urban location where they nest and roost and resource-rich foraging sites (i.e., cowshed) during the day \cite{orr_base_paper,or_r_1}. Pigeon's high density and proximity to cows during foraging highlights their potential to infect cows with a wide range of pathogens. These, in turn, may cause negative outcomes for humans, including economic losses, food shortages, and even public health threats \cite{intro_9,intro_10,intro_11}. Fig. \ref{fig:atmo} shows pigeons in a cowshed located in Israel. One can see how their spatial proximity can result in infection between the pigeons and between the pigeons and the cows.

Therefore, in this work, we propose a novel spatio-temporal multi-pathogen epidemic model for studying the spread of infectious diseases between pigeons and cows in a cowshed setting. To this end, we extended the multi-pathogen multi-species model proposed by \cite{base_paper} for our specific case and developed a detailed agent-based simulation. We used real-world data to obtain several of the model's parameter values. We hypothesize that variation in pigeons' movement (with some being more mobile or more exploratory than others) should affect disease dynamics \cite{dis_or_1,dis_or_2,dis_or_3}, and explore this effect with our model and simulation, showing a considerable effect on both magnitude and variation of outbreak indices. Furthermore, we find that if pathogen infection reduces pigeons' movement (e.g. through sickness, \cite{intro_0}) then our findings predict a positive linear correlation to the average reproduction number (ARN) of the disease.

The remaining paper is organized as follows. Section \ref{sec:model} describes the proposed model's mathematical formalization following a spatio-temporal extended SIR-based modeling approach, as well as the implementation of the proposed model as a computer simulator based on the agent-based simulation approach. Section \ref{sec:results} provides a comprehensive evaluation of the proposed model. Finally, Section \ref{sec:discussion} provides a discussion on the model's outcomes followed by suggestions for future work.  

\section{Model Definition}
\label{sec:model}
In order to capture the spatio-temporal epidemiological dynamics, we use a system of partial differential equations (PDEs). Essentially, we combine the parallel multi-pathogen with cross-species infection epidemic dynamics based on the SEIRD model \cite{spatial_2_example} together with the exploration-exploitation-based movement dynamics of the pigeons \cite{ee_intro_1,ee_intro_2,ee_intro_3}. 

\subsection{Epidemiological dynamics}
Formally, let us define a model \(M\) such that contains a finite population of pigeons (\(Pg\)) and cows (\(Cw\)) and their change over finite time \([t_0, t_f]\) such that \(t_f > t_0\), and finite space (see below). In addition, let us assume a set of disease-generating pathogens \(\Delta\) of natural size \(k \in \mathbb{N}\). At each point in time, each individual animal in the model is either susceptible (\(S\)), exposed (\(E\)), infected (\(I\)), recovered (\(R\)), or dead (\(D\)) from each of these pathogens. Thus, the epidemiological state of an animal in the model can be represented by a vector \(\eta \in \{s, e, i, r, d\}^{k}\). For instance, an individual with the state \([\{1\}, \{2\}, 0, 0, 0]\) is susceptible to the first and exposed to the second pathogen, but not infected, recovered, or dead by either of them. Therefore, each animal belongs to a super-position epidemiological state where it is susceptible, exposed, infected by, and recovered from sets of pathogens, \(s, e, i, r \subset \Delta\), such that \(s \cap e \cap i \cap r = \emptyset \wedge s \cup e \cup i \cup r = \Delta\) \cite{base_paper}. In other words, the states are pair-wise distinct and the combination of states has to include all pathogens in the model. Note that we ignore the \(d\) state since if a single pathogen caused the death of the individual, the other states \(s, e, i,\) and \(r\) are meaningless. 

\begin{figure}[!ht]
    \centering
    \includegraphics[width=0.99\textwidth]{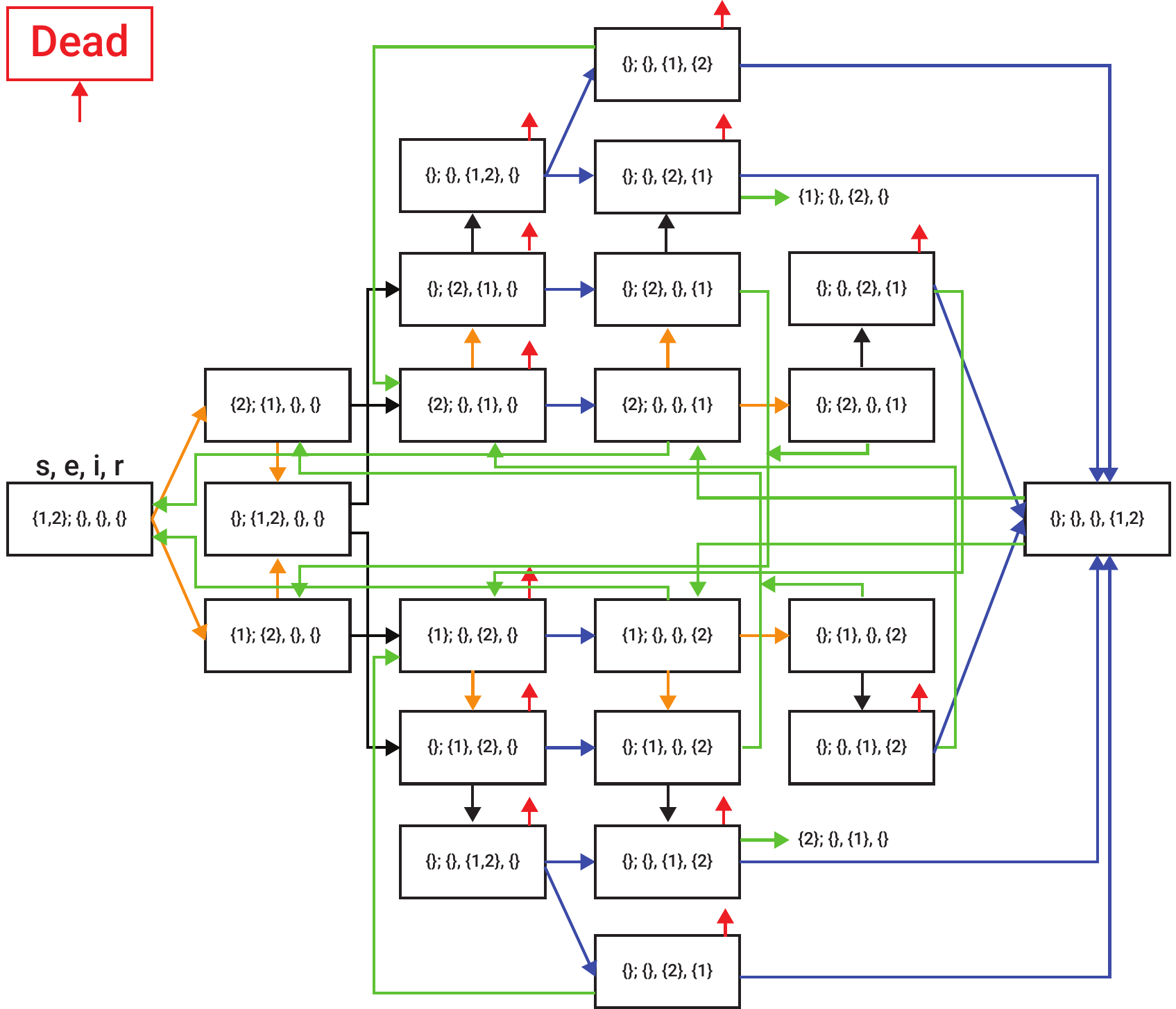}
    \caption{A schematic view of transition between disease stages, shown for \(k=2\). The red arrows indicate that from this stage, the animal might die from the disease. In a similar manner, the orange, black, blue, and green arrows indicate exposure, infection, recovery, and immunity decay, respectively. Notably, several states are duplicated for ease of reading (e.g. {2};{},{1},{}).}
    \label{fig:pandmeic_states_k_2}
\end{figure}

As such, for each state, there are 12 processes that influence the number of animals in each epidemiological state. First, animals are born at some rate \(\alpha\); Second, animals are infected by a pathogen \(j \in \Delta\) by animals from the same species, becoming exposed to it with infection rate \(\beta(x, y)\); Third, animals are infected by a pathogen \(j\) by animals from the other species, becoming exposed to it with infection rate \(\zeta(x, y)\). Forth, animals that are exposed to a pathogen \(j\) in either of the mechanisms become infectious at a rate \(\phi\); Fifth, animals infected with pathogen \(j\) recover at a rate \(\gamma\); Sixth, animals from the group \((s,e,i,r)\) are infected by a pathogen \(j \in s\) by animals from the same species, becoming exposed to it with an infection rate \(\beta(x, y)\); Seventh, animals from the group \(s,e,i,r\) are infected by a pathogen \(j \in s\) by animals from the other species, becoming exposed to it with infection rate \(\zeta(x, y)\); Eight, animals from the group \((s,e,i,r)\) which are exposed to pathogen \(j \in e\) become infectious at a rate \(\phi\); Ninth, animals from the group \((s,e,i,r)\) which are infected by pathogen \(j \in i\) recover from it at a rate \(\gamma\); Tenth, for each  \(j \in r\) animals from the group \((s,e,i,r)\) loss their immunity and become susceptible again to the pathogen \(j\) at a rate \(\psi\); Eleventh, animals from the group \((s,e,i,r)\) die due to their diseases at a rate \(\mu\); Finally, animals are naturally dying at a rate \(\upsilon\), independent of the diseases they carry (e.g. predation, trauma etc). Importantly, each parameter is also defined by a superposition of the epidemiological-subset defined by \((s,e,i,r)\). These dynamics take the partial differential equation representation as follows:

\begin{equation}
    \begin{array}{l}
    \forall s, e, i, r: \frac{\partial P_{s, e, i, r}(t, x, y)}{\partial t} = \sum_{a, b, c, d | a \cap b \cap c \cap d = \emptyset \wedge a \cup b \cup c \cup d = \Delta} \alpha_{a,b,c,d}P_{a,b,c,d} \\ \\

    + \sum_{j \in e} \beta_{s \cup j, e/j, i, r}^{s, e/j, i \cup j, r}(x, y) P_{s, e, i, r} P_{s, e, i, r}

    + \sum_{j \in e} \zeta_{s \cup j, e/j, i, r}^{s, e/j, i \cup j, r}(x, y) P_{s, e, i, r} C_{s, e, i, r}

    + \sum_{j \in i} \phi_{s, e \cup j, i/j, r} P_{s, e \cup j, i/j, r} \\ \\

    + \sum_{j \in r} \gamma_{s, e, i \cup j, r/j} P_{s, e, i \cup j, r/j} 

    + \sum_{j \in s} \psi_{s/j, e, i, r \cup j} P_{s/j, e, i, r \cup j} 

    - \sum_{j \in s} \beta_{s, e, i, r}^{s, e/j, i \cup j, r}(x, y) P_{s, e, i, r} P_{s, e/j, i \cup j, r}  \\ \\ 

    - \sum_{j \in s} \zeta_{s, e, i, r}^{s, e/j, i \cup j, r}(x, y) P_{s, e, i, r} C_{s, e/j, i \cup j, r} 

    - \sum_{j \in e} \phi_{s, e, i, r} P_{s, e, i, r}

    - \sum_{j \in r} \psi_{s, e, i, r} P_{s, e, i, r} \\\\

    - \sum_{j \in i} \gamma_{s, e, i, r} P_{s, e, i, r} - \sum_{j \in i}  \mu_{s, e, i, r} P_{s, e, i, r} - \upsilon_{s, e, i, r} P_{s, e, i, r}
    
    \end{array} 
    \label{eq:pandemic_temporal}
\end{equation}
Similarly, the cows' epidemiological dynamics are identical to the pigeon's one while having different parameter values. A schematic view of the epidemiological states of the model for the specific case of two pathogens (i.e., \(k=2\)) is shown in Fig. \ref{fig:pandmeic_states_k_2} where each box indicates the epidemiological state of the individual represented by the pathogens belongs to each of the \(s, e, i, r\) sets. For example, a healthy individual starts on the left box and moves to the right as it is exposed to a pathogen (following an orange arrow), becoming infectious (following a black arrow), recovers (following a blue arrow), and finally return to be again susceptible due to immunity decay (following a green arrow). During the infectious state, there is a chance the individual would die due to the pathogen as indicated by the red arrow, as well as background natural mortality throughout its life (not shown). 

\subsection{Movement dynamics}
\begin{figure}[!ht]
    \centering
    \includegraphics[width=0.99\textwidth]{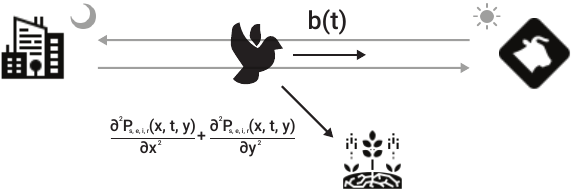}
    \caption{A schematic view of a pigeon movement in a 2-dimensional space with the exploitation vector (\(b(t)\)) at a specific time and over time (marked by the gray line) and the exploration vector (\(\frac{\partial^2 P_{s,e,i,r}(t,x,y)}{\partial x^2} + \frac{\partial^2 P_{s,e,i,r}(t,x,y)}{\partial t^2}\)) which looking for resources (such as food). For our model, all these alternative explored sites can be condensed into a single one as they do not directly affect transmission dynamics.}
    \label{fig:scheme_spatial}
\end{figure}

The movement dynamics, unlike the epidemiological dynamics, are unique for each species, reflecting their different life histories and wild/captive contexts. First, as cows are living in a relatively small cowshed and interact intensively with each other, it can be approximated that they are well-mixing within each farm \cite{first_sir}. This decision was motivated by the infection range compared to the animals' proximity as indicated in Fig. \ref{fig:atmo_1}. Namely, the probability that a cow meets any other cow in the population for any point in time is identical and proportional to the cow population size. As such, we assume the cow population does not have any movement dynamic which is significant for the proposed model. On the other hand, feral pigeons worldwide (and our system included) are known to sleep in buildings while traveling to forage in food sites during the day \cite{move_bio}. However, pigeons explore their surroundings and alternate among different foraging sites, reflecting a tradeoff between the exploitation of known resources and the exploration of new ones. Individuals differ in their tendency to explore \cite{pop_orr_1}, with some visiting sites of interest (cowsheds) more frequently than others \cite{fly_pattern}. Indeed, previous studies show that one can explain the pigeon flight patterns using the exploration-exploitation model \cite{peigon_fly_1,peigon_fly_2,peigon_fly_3}. Moreover, as pigeons are exposed and infected with pathogens, their flight abilities might be reduced \cite{sick_fly_less}. Hence, the pigeons' movement dynamics can be represented as a weighted average of a random walk (representing the exploration dynamics), and a time-based directed walk (which represents the exploitation dynamics). Moreover, both behaviors are influenced by the combination of the pathogens each pigeon is exposed to and infected with. As such, the pigeons' movement dynamics take the form:
\begin{equation}
    \forall e, i: \frac{\partial P_{s,e,i,r}(t,x,y)}{\partial t} =  \omega_1^{e, i} \big ( \frac{\partial^2 P_{s,e,i,r}(t,x,y)}{\partial x^2} + \frac{\partial^2 P_{s,e,i,r}(t,x,y)}{\partial t^2} \big ) + \omega_2^{e, i} b_{s,e,i,r}(t),
    \label{eq:pandemic_spatial}
\end{equation} 

where \(\omega_1^{e, i}, \omega_2^{e, i} \in \mathbb{R}^+\) are the coefficients of movement such that \(\omega_1^{e, i}/\omega_2^{e, i}\) is the exploration-exploitation rate and \(\omega_1^{e, i} + \omega_2^{e, i}\) is the total amount of movement for a single step in time. In addition, \(b_{s,e,i,r}(t)\) is the average time-depended directed walk vector of the pigeons' population and satisfies that \(\forall s, e, i, r, t: |b_{s, e, i, r}(t)| = 1\). Fig. \ref{fig:scheme_spatial} presents a schematic view of the movement dynamics. 

\subsection{Agent-based Simulation}
\label{sec:simulation}
Since the proposed model (see Eqs. (\ref{eq:pandemic_temporal} and \ref{eq:pandemic_spatial}) captures population dynamics, it is, in practice, takes into consideration only the average behavior of the population \cite{spatial_2_example,economic_2}. Another shortcoming of such SIERD model is that practical applications will be limited by the availability of required parametrization data. Thus, inspired by \cite{teddy_sir_review}, we implemented the proposed model using the agent-based simulation (ABS) approach. Generally speaking, ABS is able to provide another layer of realism by allowing each animal in the population with unique attributes, reflecting more closely capture the realistic dynamics (in contrast to the system at equilibrium) observed in nature where animals within a population differ \cite{pop_orr_1,pop_orr_2}. Moreover, ABS makes the infection computation burden relatively small, as one can use spatial approximation to determine interactions between individuals in the population(s), and it allows exploring the importance of specific parameter values of general scenarios for emerging population-level patterns. 

Formally, let \(M := (Pg, Cw)\) be a tuple of agents' sets which moves and interacts in discrete finite time steps \(t \in [1, \dots, T]\), where \(T<\infty\). In order to use the ABS approach, one has to define the agents in the dynamics as well as their three types of interactions: agent-agent, agent-environment, and spontaneous (i.e., depends only on the agent's state and time) \cite{agent_based_exp_1}. To this end, for our model, each agent \(a\) is represented by a finite state machine \cite{fsm} as follows: \(a := (\xi, x, y, \eta, \{w_1\}^{e, i}, \{w_2\}^{e, i}, \beta_r)\) where \(\xi \in \{p, c\}\) is the agent's species (i.e., pigeon or cow), \(x, y\) are the spatial coordinates of the agents in a Cartesian coordinate system, \(\eta\) is the agent's epidemiological state, \(\forall e, i \forall j \in \{1, 2\}: \{w_1\}^{e, i}\) are the agent's personal spatial exploration-exploitation coefficients, and \(\beta_r\) is a vector representing infection radius of each pathogen, effectively setting the relevant interaction distance among agents. 

At the first time step (\(t=1\)), the pigeons and cow populations (\(P, C\)) are generated based on some initial conditions and located on a continuous two-dimensional map with dimensions denoted by \(w, h \in \mathbb{R}^+\). Then, at each time step \(t\), each individual in the pigeon follows Eq. (\ref{eq:pandemic_spatial}). The decisions of all individuals are referred to as the pigeon population walk. Following standard convention, we assume that all individuals may be located at each point on the map. Given the nature of our particular system, this general simplification is very suitable as pigeons aggregate at high numbers in high proximity (centimeters) with a movement range of (kilometers). For example, Fig. \ref{fig:atmo} shows this proximity. Between every two consecutive time steps, individuals infected each other based on their proximity, here simplified to a threshold function (note that we simplify these to a single value rather than a pathogen-specific effective distance). Namely, if an agent is infected with pathogen \(j\) has another agent susceptible to pathogen \(j\) and is located within a range of \(\beta_r^j\), it has a probability of \(\beta \in [0, 1]\)  for the specific \(\eta = (s, e, i, r)\) state of the infectious and susceptible individuals if they are from the same species and with probability \(\zeta \in [0, 1]\), otherwise. Thus, applying the epidemiological dynamics represented in Eq. (\ref{eq:pandemic_temporal}) in a spatially-local manner. In addition, the spontaneous epidemiological process of exposure to infectious, recovery, death, and immunity-decay are computed using a rate associated with the number of time steps rather than being computed using the population-level dynamics as suggested by Eq. (\ref{eq:pandemic_temporal}), following a common practice of ABS implementation \cite{software_hard_3}. For instance, an individual exposed to pathogen \(i\) would become infectious to this pathogen after \(\psi_i\) time steps. Lastly, we compute the metrics and store them. The simulation ends either after \(T\) steps in time or once all individuals in the model are dead. 

\section{Experiments}
\label{sec:results}
In this section, we perform \textit{in silico} experiments based on the proposed model. First, we find from the literature realistic values ranges for the model's parameters to obtain biologically relevant instances of the proposed model. Afterwards, using these instances we explore the central spatial and temporal dynamics occurring in such a system. 

\subsection{Setup}
While high-resolution and extensive epidemiological data required to obtain a real-world instance of the proposed model is currently unavailable (to our best knowledge), partial data in the form of the pigeons' movement dynamics and some biological data about pathogens' spread are available in the literature. Specifically, prior empirical work has shown that pigeons in general and in our system in particular carry a diversity of pathogens simultaneously. Hence, we used the data collected by \cite{orr_base_paper}. Namely, \cite{orr_base_paper} captured \(n=328\) pigeons from three cowsheds located in Israel. The authors installed a GPS device on several of those and were able to obtain sufficient tracking data from  \(n=33\) individuals. providing its location in \(3\) meters accuracy every \(10\) minutes. The transmitters were active for \(214 \pm 193\) days (range: 14 and 713 days per individual). In total, they collected \(8635\) tracking days regarding the location of the pigeons over time. In addition, all three hundred pigeons were sampled with oral and cloacal swabs for pathogen identification. a subset of 29 samples was also assessed for microbiome-wide DNA presence using next-generation sequencing \cite{next_gen_dna}. Table \ref{table:parameters} summarizes the proposed model's hyper-parameter space values for each configuration based on this study and values from the biological literature \cite{virus_intro_1,virus_intro_2,virus_intro_3,pip_5,hard_msms_1,virus_1,virus_2,virus_3}. In particular, in agreement with the above GPS sampling rate, we chose to simulate a 10-minute time step, balancing the computational burden and the model's accuracy. Namely, the movement dynamics as well as the infection can be approximated in a several minutes scale \cite{orr_base_paper,spatio_temporal_airborne_andemic_ariel} and the other dynamics are much slower. Moreover, the population sizes are chosen based on the estimation of three cowshed workers. The exploration-exploitation coefficients are computed based on the average behavior from \cite{orr_base_paper}.

\begin{table}[!ht]
\centering
\begin{tabular}{p{0.1\textwidth}p{0.45\textwidth}p{0.35\textwidth}}
\hline \hline
\textbf{Symbol} & \textbf{Description} & \textbf{Default value} \\ \hline \hline
   \(T\) & Number of simulation rounds (spanning over a year in the \(\Delta t\) used)  & 51840 \\ 
   \(\Delta t\) & Simulation round's duration in time & 10 minutes \\ 
   \(|P(0)|\) & The initial pigeons' population size & [500, 5000]   \\
   \(|C(0)|\) & The initial cows' population size & [100, 1000] \\
   \(k\)   & The number of pathogens & \([2-7]\)  \\
   \((w, h)\) & The two-dimensional spatial map dimentions &  \(1-10 \times 1-10\) kilometers  \\
   \(\alpha\) & Birth rate  & Cows: 0, Pigeons: \(1.8 \cdot 10^{-3}\)  \\
   \(\beta\) & Same species infection rate  & \([1.5 \cdot 10^{-5}, 9.5 \cdot 10^{-5}]\)  \\
   \(\zeta\) & Cross-species infection rate & \([1.5 \cdot 10^{-5}, 9.5 \cdot 10^{-5}]\) \\
   \(\phi\) & Exposure to infectious transformation rate  & \([6.9 \cdot 10^{-3}, 6.9 \cdot 10^{-4}]\) \\
   \(\gamma\) & Recovery rate  & \([1.7 \cdot 10^{-4}, 7.8 \cdot 10^{-3}]\) \\
   \(\psi\) & Immunity decay rate  & \([5 \cdot 10^{-6}, 10^{-4}]\)  \\
   \(\upsilon\) & Natural death rate & Cows: \(3.86 \cdot 10^{-7}\), Pigeons: \(1.48 \cdot 10^{-5}\) \\
   \(\omega_1, \omega_2\) & Exploration-exploitation walk coefficients & 0.16, 2.3 meter/minute  \\
   \(\beta_r\) & Spatial infection radius & 5 meters  \\ \hline
\end{tabular}
\caption{A summary of the proposed model's parameters and hyperparameters with their realistic and theoretical value ranges. }
\label{table:parameters}
\end{table}

\begin{figure}[!ht]
    \centering
    \includegraphics[width=0.66\textwidth]{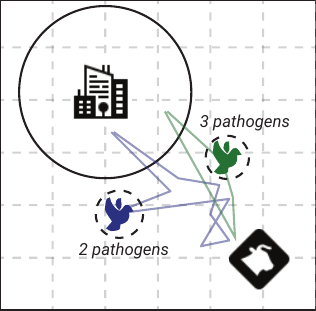}
    \caption{A schematic view of the synthetic scenario generation process.}
    \label{fig:synt_design}
\end{figure}

In order to investigate the epidemic spread dynamics from various scenarios, one first needs to define the setup of the model. Hence, we uniformly randomly sample the model's parameter values from Table \ref{table:parameters}. For the hyperparameters, we uniformly randomly sample \(|C(0)|\) to be between 100 and 1000, \((|P(0)|\) to be between 500 and 5000, the number of pathogens \(k\) to be between 2 and 7, and the map's size to be between 1 and 10 kilometers. For simplicity, we positioned the cowshed in the center of the map and placed the urban locations in which pigeons nest at a random point in the map which is at least half a kilometer from the cowshed (well within pigeons' daily movement range but two orders of magnitude larger than the direct transmission threshold (\(\beta_r\)). Then, each pigeon is located in a personal nesting location which is normally distributed around the center location of the urban area with a mean and standard deviation of \(0.25 \pm 0.1\) kilometers (effectively implying most pigeons will be nesting inside the urban area). In addition, we set the simulation duration to be  \(51,840\) time steps of ten minutes to obtain a total of one year. Fig. \ref{fig:synt_design} shows a schematic view of the synthetic scenario generation process. This setup is also repeated with multiple cowsheds such that the cowsheds are too far apart to allow cows to infect each other between cowsheds but close enough for the same pigeons to visit all cowsheds in the region. 

In order to evaluate the epidemic spread, one is required to define an epidemiological metric of interest. Here we consider three of the most popular epidemiological metrics: the ARN (\(E[R_t]\)), the maximum number of infections (MI), and the number of dead cows (CD) due to the epidemic \cite{metric_paper_1,metric_paper_4,first_teddy_paper,metric_paper_2,metric_paper_3}. \(R_t\) measures the number of secondarily infected individuals given the epidemic state at a given time \(t\) \cite{metric_paper_3}. Intuitively, the ARN (\(E[R_0]\)) computes how many, on average, a single infected individual infects other individuals. The max infected metric at a time point in time, \(t\), counts the number of individuals infected by some pathogen divided by the population size. The cow death computes how many cows are dead up to some point in time, \(t\). Formally, \(R_t\) can be approximated using the following formula: \(R_t := \big ( I(t) - I(t-1) + R(t) - R(t-1)  \big ) / I(t-1)\), the \(MI\) at time \(t\) is defined as follows \(MI(t) := \max_{i \in [t_0, t_t]} I(i)\), and \(CD\) at time \(t\) is defined to be \(\sum_{s, e, i, r \in \Delta} \mu_{s, e, i, r} C_{s, e, i, r}\), such that \(I(t) := \|\{\forall p \in P | \eta[i]_t \neq \emptyset \}\|\) and \(R(t) := \|\{\forall p \in P | \eta[r]_t \neq \emptyset \}\|\) where \(\eta[x]_t\) indicates the set of pathogens for the \(x \in \{s, e, i, r, d\}\) epidemiological state at time \(t\). For our case, we assume that both \(R_t\) and \(MI\) are computed only for the pigeon population while \(CD\) naturally considers the cow population. 

\subsection{Results}
\begin{figure}[!ht]
    \centering
    \begin{subfigure}{.32\textwidth}
        \includegraphics[width=0.99\textwidth]{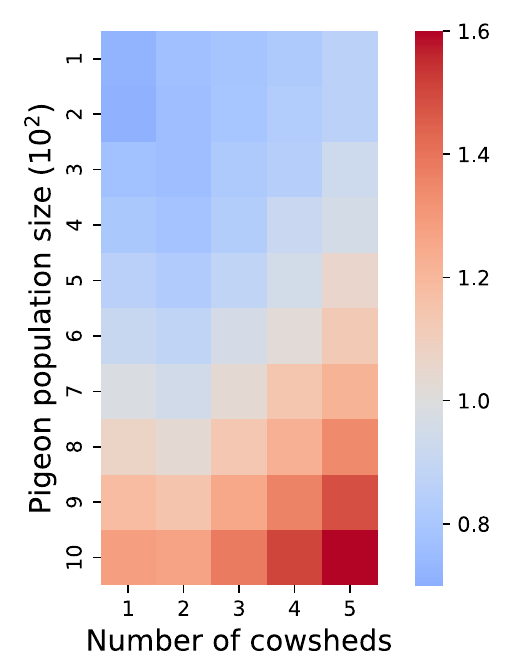}
        \caption{ARN (\(E[R_0]\)).}
        \label{fig:cowsheds_r_zero}
    \end{subfigure}
    \begin{subfigure}{.32\textwidth}
        \includegraphics[width=0.99\textwidth]{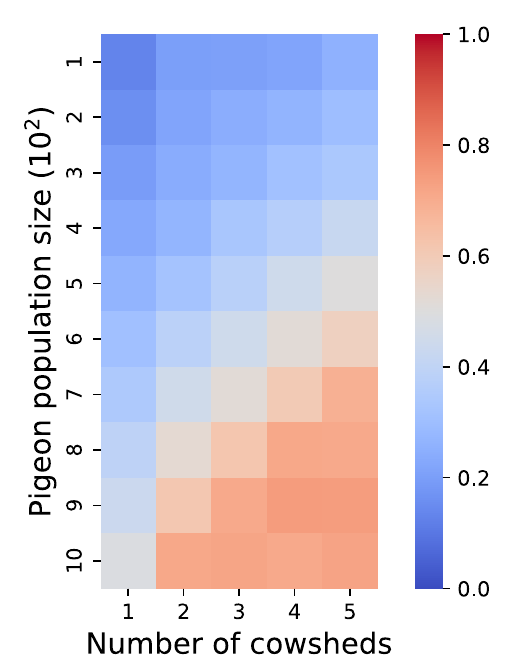}
        \caption{Max infected (MI).}
        \label{fig:cowsheds_mi}
    \end{subfigure}
    \begin{subfigure}{.32\textwidth}
        \includegraphics[width=0.99\textwidth]{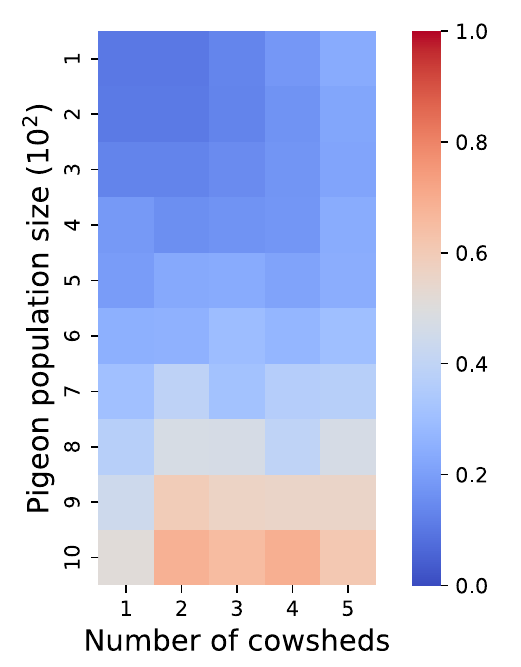}
        \caption{Cow death portion (CD).}
        \label{fig:cowsheds_cd}
    \end{subfigure}
    
    \caption{Epidemic metrics as a function of the pigeon population size and number of cowsheds with 50 cows at each one. The results are shown as a mean of \(n=1000\) simulation realizations. Notably, all three metrics show an increase with a growing number of pigeons, in interaction with the number of cowsheds}
    \label{fig:cowsheds}
\end{figure}

Based on this setup, we conducted three main experiments as well as a sensitivity analysis for the model. First, let us consider a scenario in which the number of cowsheds ranges from the basic case of a single cowshed to a more complex one of up to five spatially distinct cowsheds. As the number of parameters and their values can range widely from one instance of the model to another, we computed the model for \(n = 1000\) simulation realizations, each time with a random sample of the models' parameters' values using a uniform distribution. Fig. \ref{fig:cowsheds} presents the epidemic metrics as a function of the pigeon population size and number of cowsheds with 50 cows at each one. The results are shown as a mean of \(n=1000\) simulation realizations. Intuitively, one can notice that as the pigeons' population size increases (lower in the y-axis), all three epidemic metrics also increase, on average. Moreover, a phase transition between a single and multiple cowshed(s) is revealed. This phase transition can be associated with the fact that multiple cowsheds cannot infect each other without the transition of pigeons which is less likely for a smaller pigeons population size. 

\begin{figure}[!ht]
    \centering
    \begin{subfigure}{.32\textwidth}
        \includegraphics[width=0.99\textwidth]{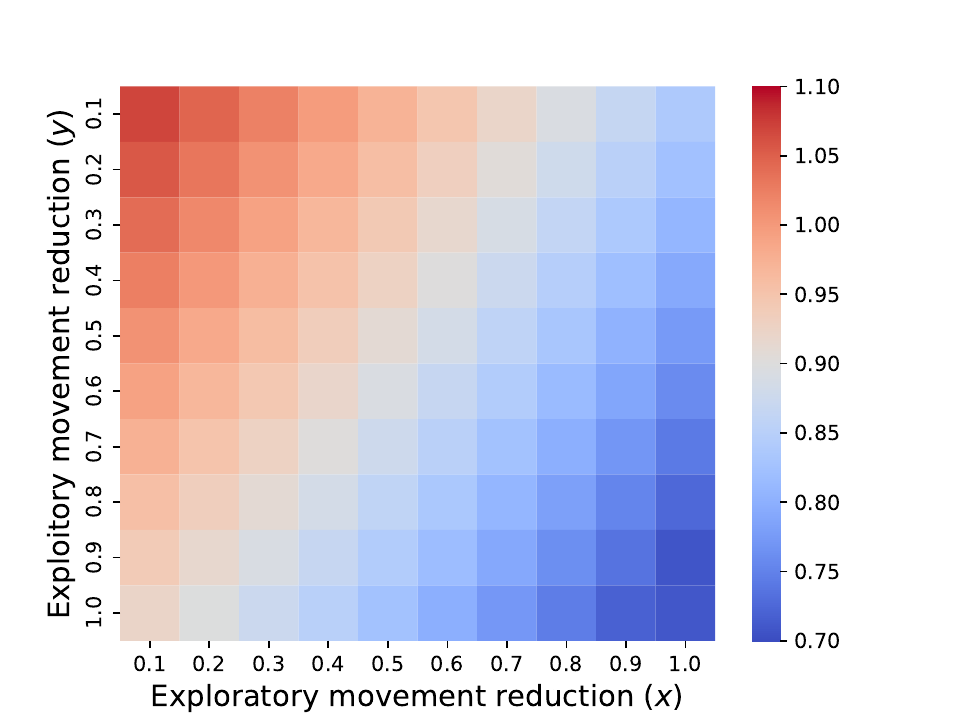}
        \caption{ARN (\(E[R_0]\)).}
        \label{fig:sick_movement_r_zero}
    \end{subfigure}
    \begin{subfigure}{.32\textwidth}
        \includegraphics[width=0.99\textwidth]{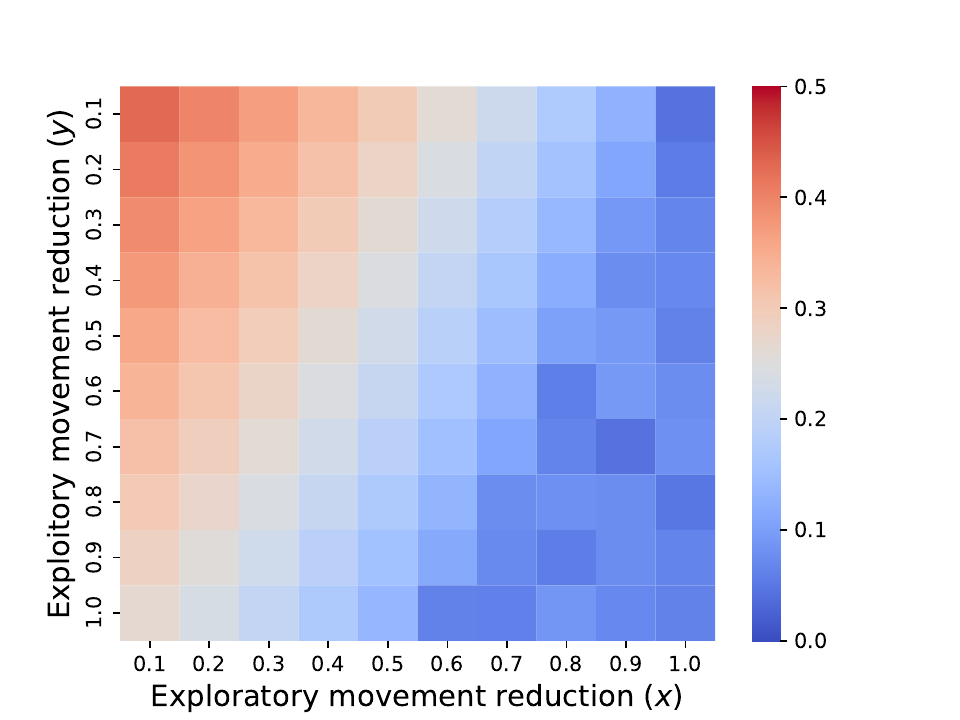}
        \caption{Max infected (MI).}
        \label{fig:sick_movement_mi}
    \end{subfigure}
    \begin{subfigure}{.32\textwidth}
        \includegraphics[width=0.99\textwidth]{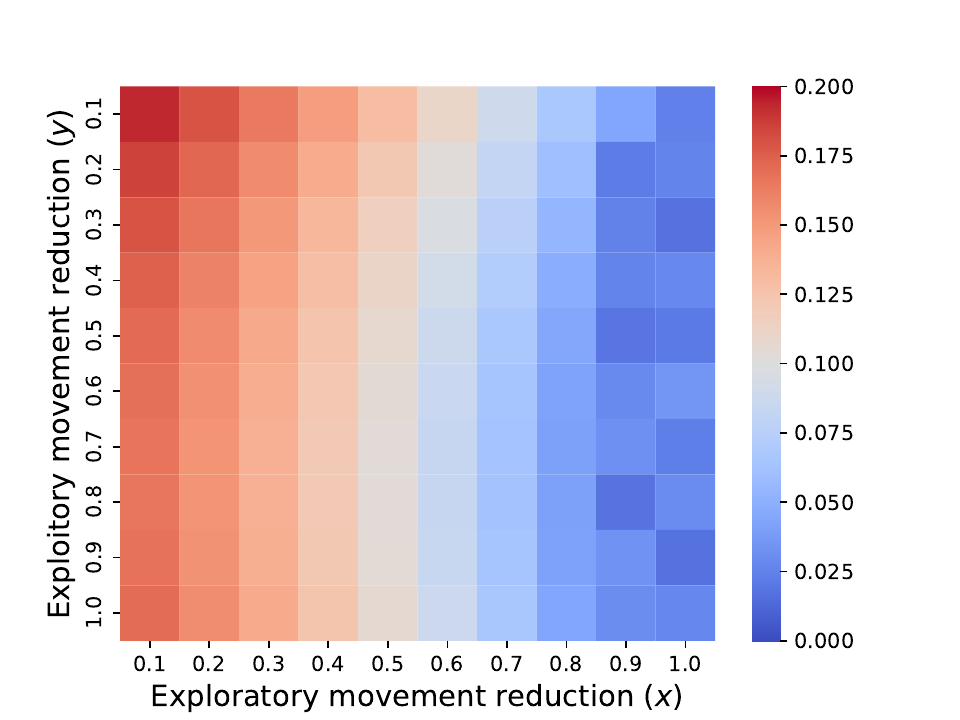}
        \caption{Cow death portion (CD).}
        \label{fig:sick_movement_cd}
    \end{subfigure}
    
    \caption{Epidemic metrics as a function of the pigeon movement reduction. The results are shown as a mean of \(n=1000\) simulation realizations (representing a random sample from the hyper-parameter space). For all three epidemiological metrics, a growing reduction in exploratory mobility (jumps between sites) causes a reduction in the pandemic spread, with a stronger impact compared to the reduction in pigeons' exploitation tendency (re-visits to a site). }
    \label{fig:sick_movement}
\end{figure}

In a similar manner, since pigeons operate as infectious vectors, they are commonly sick with one or even many pathogens in parallel (here only two are included in the simulation) which might reduce their movement. Hence, let us consider a simplistic but biologically-supported scenario  \cite{peigon_fly_1,peigon_fly_3} where each pathogen, regardless of its nature, experiences a reduction of \(x \in [0, 1]\) and \(y \in [0, 1]\) in the exploration-exploitation (\(w_1, w_2\)) parameters of an infected pigeon, respectively. Fig. \ref{fig:sick_movement} presents the epidemic metrics as a function of the pigeon movement reduction. Namely, a reduction of \(x=0.5\) implies that a sick pigeon moved with \(\omega_1 \cdots x = \omega_1 \cdot 0.5\)  and y=0.5 implies \(\omega_2 \cdots y = \omega_2 \cdot 0.5\). The results are shown as a mean of \(n=1000\) simulation realizations. Unsurprisingly, as \(x\) and \(y\) increase, all three epidemiological metrics decrease, since effectively pigeons proved lower connectivity. The decrease is quicker than the decrease rate of \(x\) while negatively linear to the decrease of \(y\). In other words, the exploitation tendency has a weaker effect on the pandemic spread compared to the exploration one, which indicates that the number of cowshed visits a pigeon performs, on average, does not have much effect as it will transmit the disease to cows anyway.

\begin{figure}[!ht]
    \centering
    \begin{subfigure}{.32\textwidth}
        \includegraphics[width=0.99\textwidth]{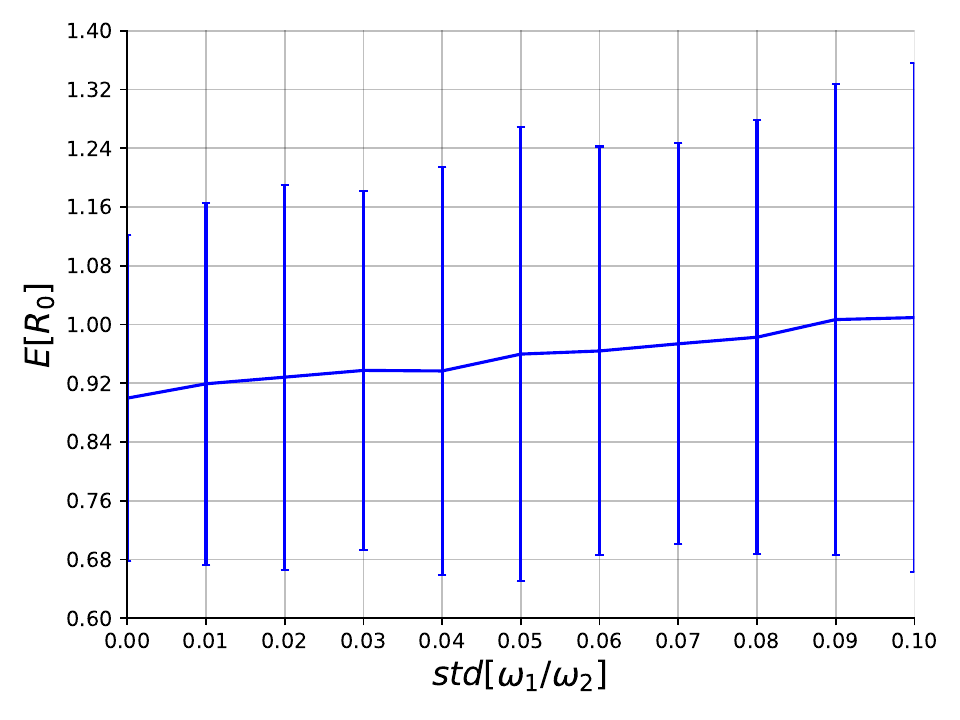}
        \caption{ARN (\(E[R_0]\)).}
        \label{fig:r_zero}
    \end{subfigure}
    \begin{subfigure}{.32\textwidth}
        \includegraphics[width=0.99\textwidth]{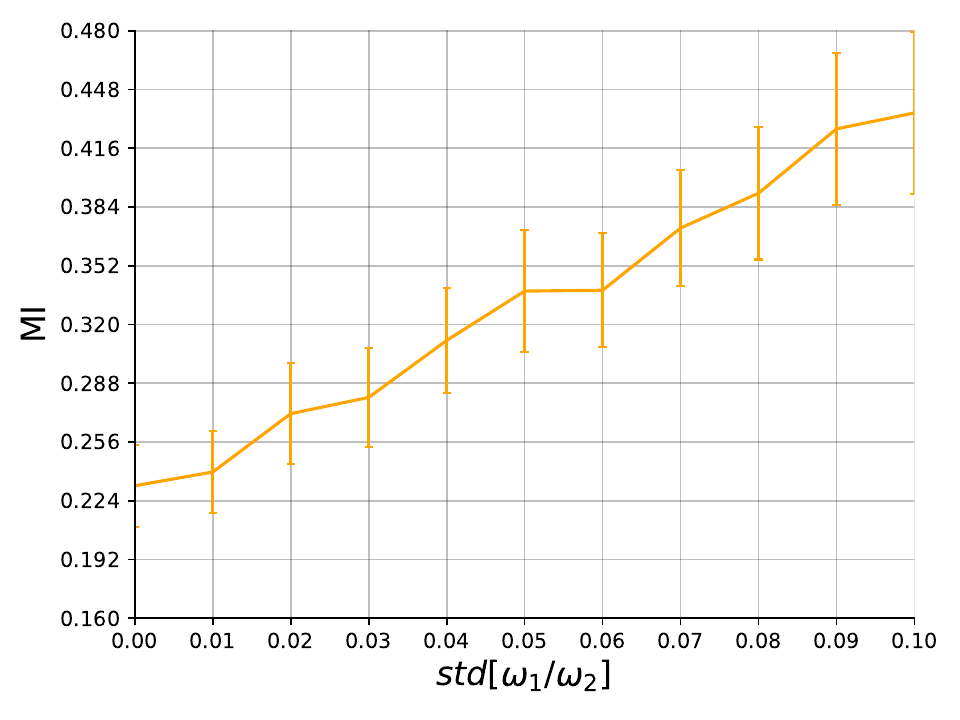}
        \caption{Max infected (MI).}
        \label{fig:mi}
    \end{subfigure}
    \begin{subfigure}{.32\textwidth}
        \includegraphics[width=0.99\textwidth]{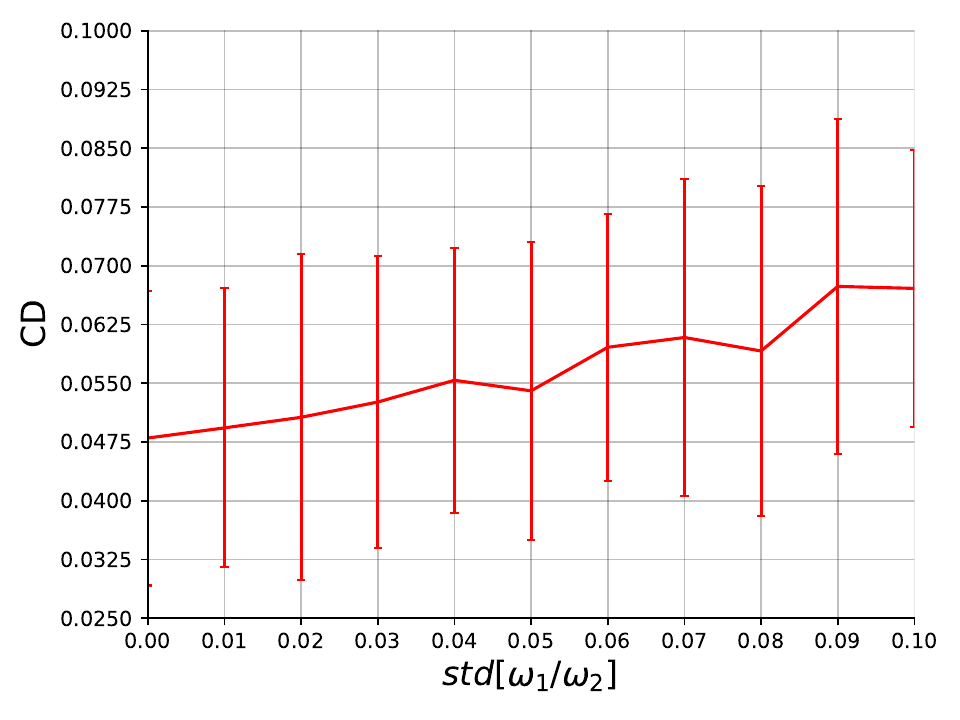}
        \caption{Cow death portion (CD).}
        \label{fig:cd}
    \end{subfigure}
    
    \caption{A heterogeneous analysis of the pigeons' population movement. The results are shown as the mean \(\pm\) standard deviation value of \(n = 1000\) simulation realizations. Notably, as \(std[\omega_1/\omega_2]\) increase, all three epidemiological metrics also increase and their standard deviation, as indicated by the error bars, is also increased.}
    \label{fig:last_fig}
\end{figure}

To further investigate the influence of the heterogeneous movement between pigeons, we used the values of \(\omega_1\) and \(\omega_2\) from Table \ref{table:parameters} but changed the standard deviation of \(\omega_1/\omega_2\) (denoted by \(std[\omega_1/\omega_2]\)) using the ratio of  \(\omega_1/\omega_2 = 6.95 \cdot 10^{-3}\) as the reference value (this ratio represent the ratio between the two empirically observed values from \cite{orr_base_paper}). Intuitively, a larger standard deviation in the exploitation-exploration rate indicates more diversity in the movement dynamics in the population and therefore more heterogeneously. Fig. \ref{fig:last_fig} summarizes the results of this analysis, showing the values as the mean \(\pm\) standard deviation outcome of \(n = 1000\) simulation realizations. Notably, as \(std[\omega_1/\omega_2]\) increase, all three epidemiological metrics also increase both in mean value and in their standard deviation, as indicated by the error bars. This result indicates that behavioral variation among hosts can affect system-level stability and result in both more pronounced outbreaks (e.g. ARN increases from 0.92 to 1 for this particular value of \(\omega_1/\omega_2\) ), and in higher variation among realizations, reflecting higher sensitivity to the specific parameters of each one.

\begin{figure}[!ht]
    \centering
    \begin{subfigure}{.32\textwidth}
        \includegraphics[width=0.99\textwidth]{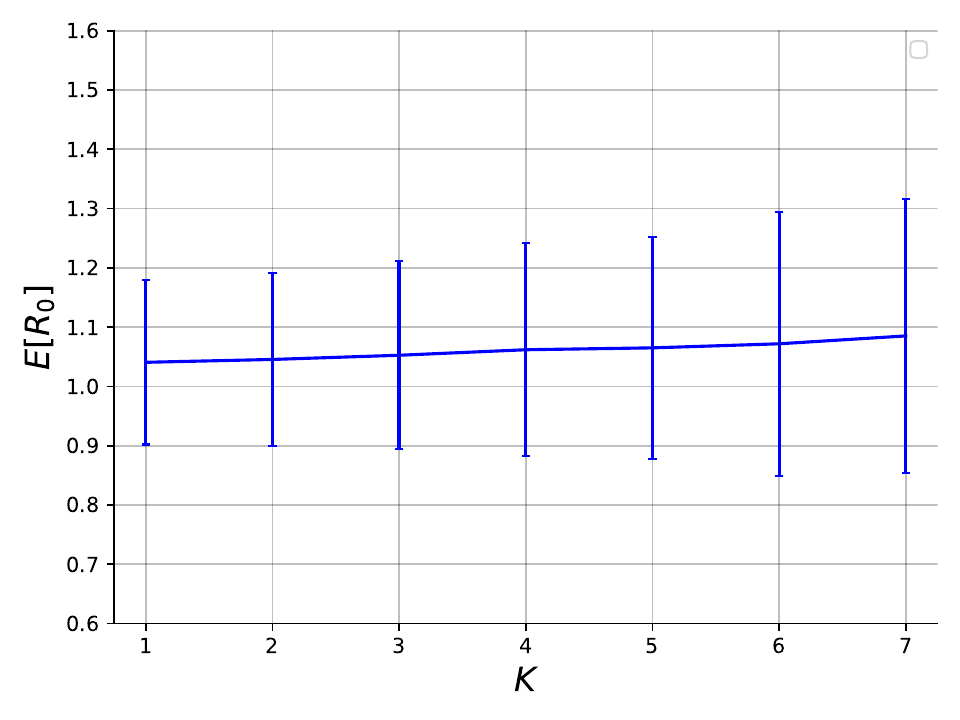}
        \caption{Pathogen count (\(k\)) - \(E[R_0]\).}
        \label{fig:sens_k_e_r_0}
    \end{subfigure}
    \begin{subfigure}{.32\textwidth}
        \includegraphics[width=0.99\textwidth]{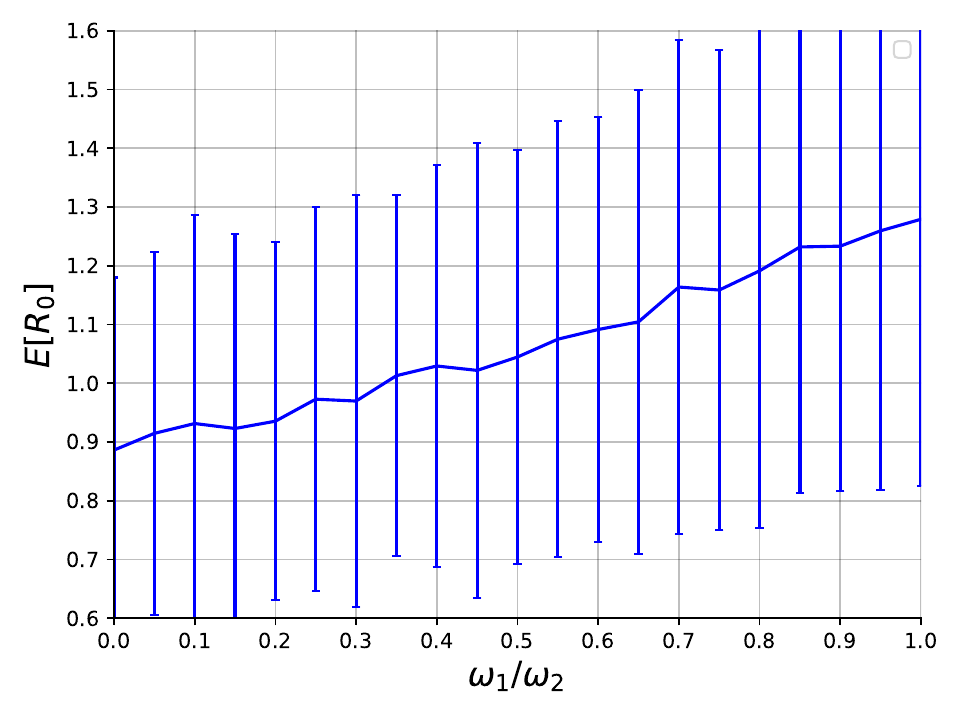}
        \caption{EE walk ratio  (\(\omega_1/\omega_2\)) - \(E[R_0]\).}
        \label{fig:sens_omegas_e_r_0}
    \end{subfigure}
    \begin{subfigure}{.32\textwidth}
        \includegraphics[width=0.99\textwidth]{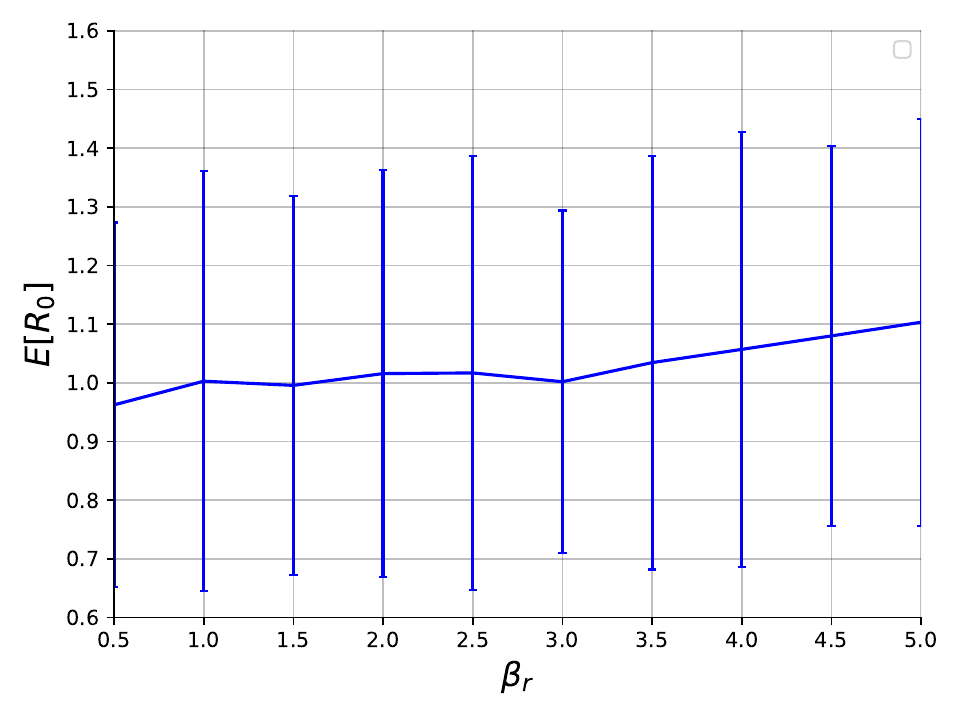}
        \caption{Spatial infection radius (\(\beta_r\)) - \(E[R_0]\).}
        \label{fig:sens_beta_e_r_0}
    \end{subfigure}

    \begin{subfigure}{.32\textwidth}
        \includegraphics[width=0.99\textwidth]{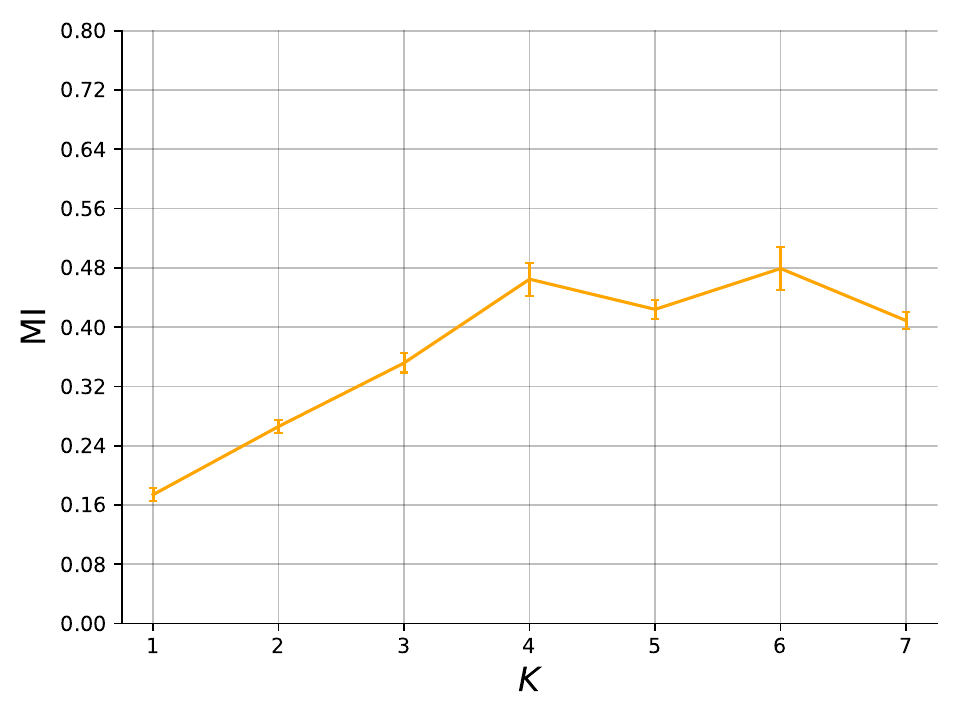}
        \caption{Pathogen count (\(k\)) - \(MI\).}
        \label{fig:sens_k_mi}
    \end{subfigure}
    \begin{subfigure}{.32\textwidth}
        \includegraphics[width=0.99\textwidth]{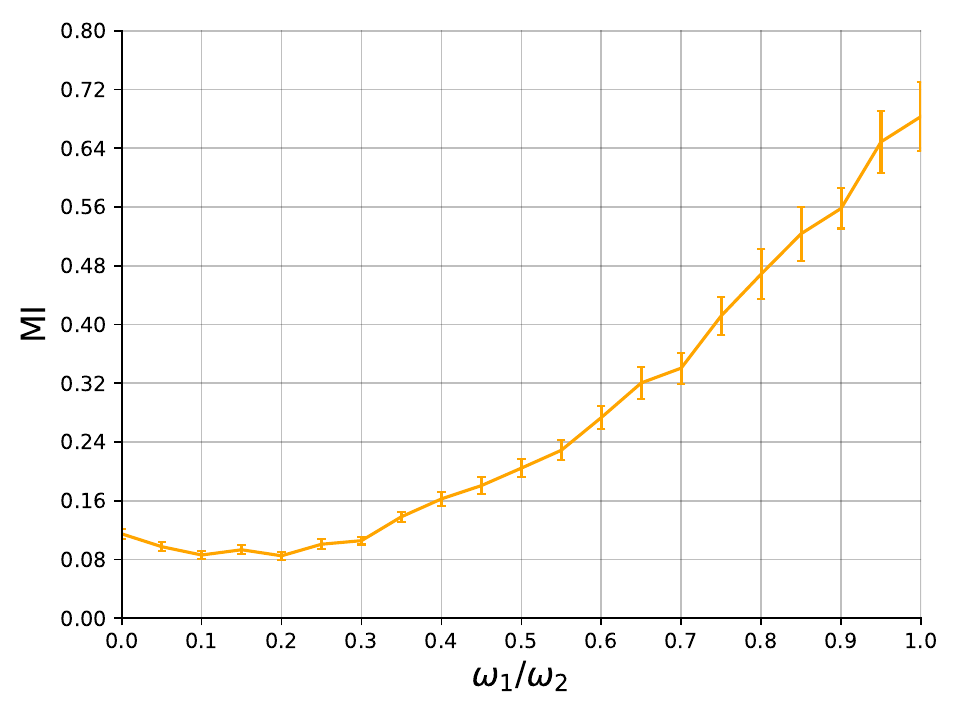}
        \caption{EE walk ratio  (\(\omega_1/\omega_2\)) - \(MI\).}
        \label{fig:sens_omegas_mi}
    \end{subfigure}
    \begin{subfigure}{.32\textwidth}
        \includegraphics[width=0.99\textwidth]{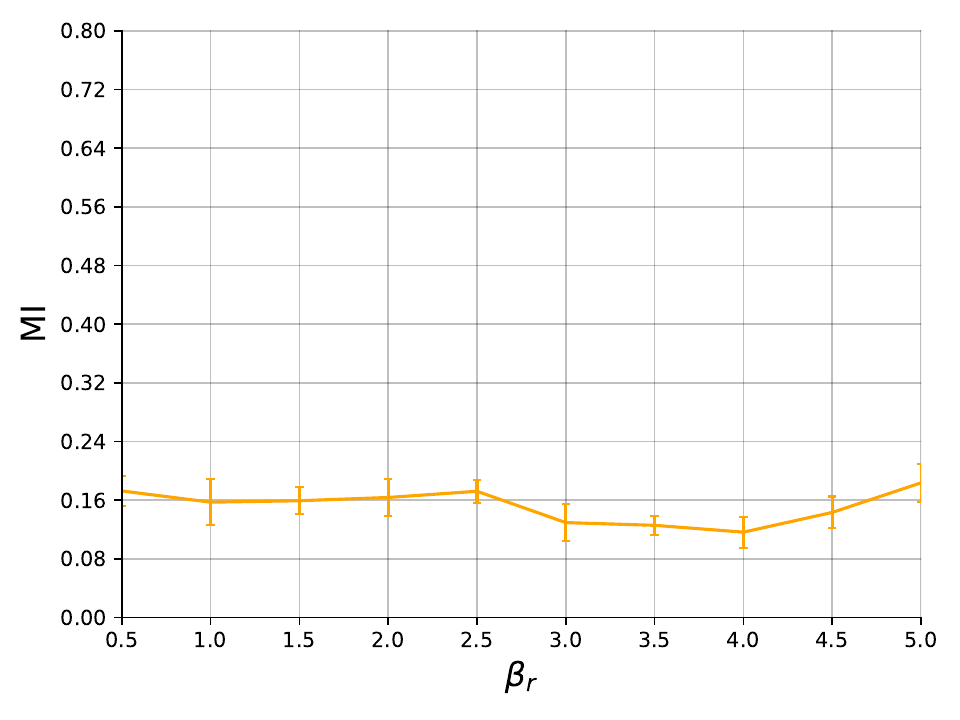}
        \caption{Spatial infection radius (\(\beta_r\)) - \(MI\).}
        \label{fig:sens_beta_mi}
    \end{subfigure}

    \begin{subfigure}{.32\textwidth}
        \includegraphics[width=0.99\textwidth]{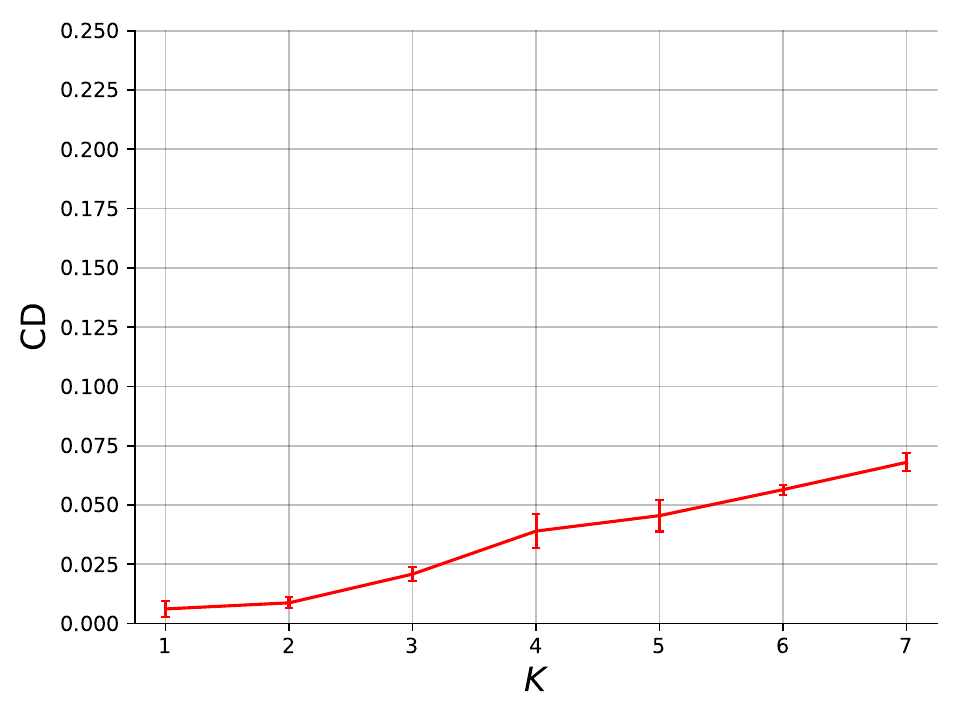}
        \caption{Pathogen count (\(k\)) - \(CD\).}
        \label{fig:sens_k_cd}
    \end{subfigure}
    \begin{subfigure}{.32\textwidth}
        \includegraphics[width=0.99\textwidth]{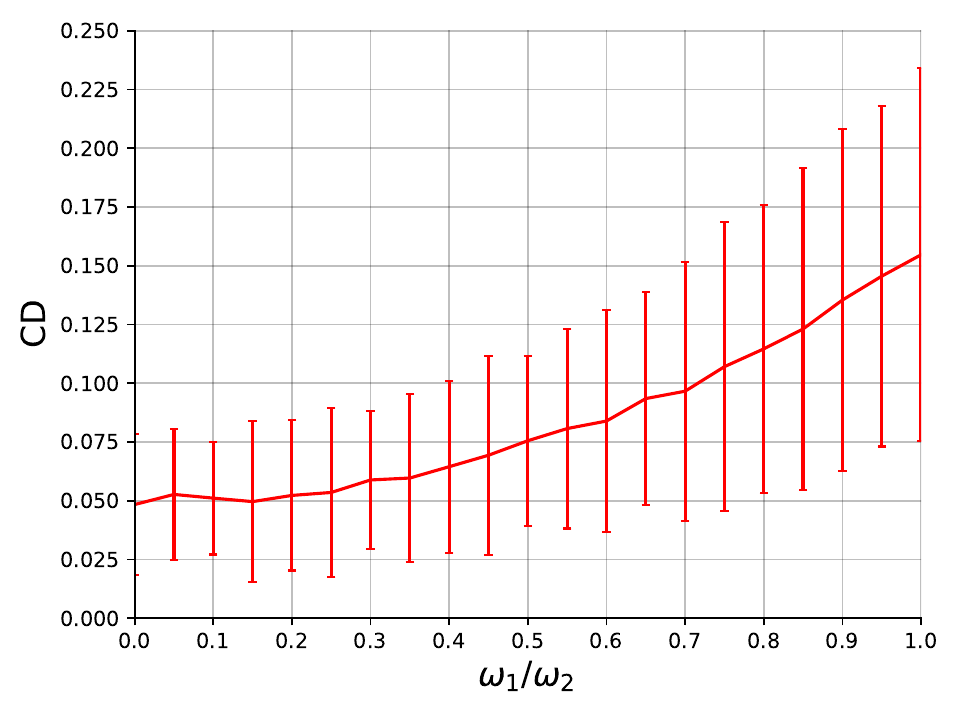}
        \caption{EE walk ratio  (\(\omega_1/\omega_2\)) - \(CD\).}
        \label{fig:sens_omegas_cd}
    \end{subfigure}
    \begin{subfigure}{.32\textwidth}
        \includegraphics[width=0.99\textwidth]{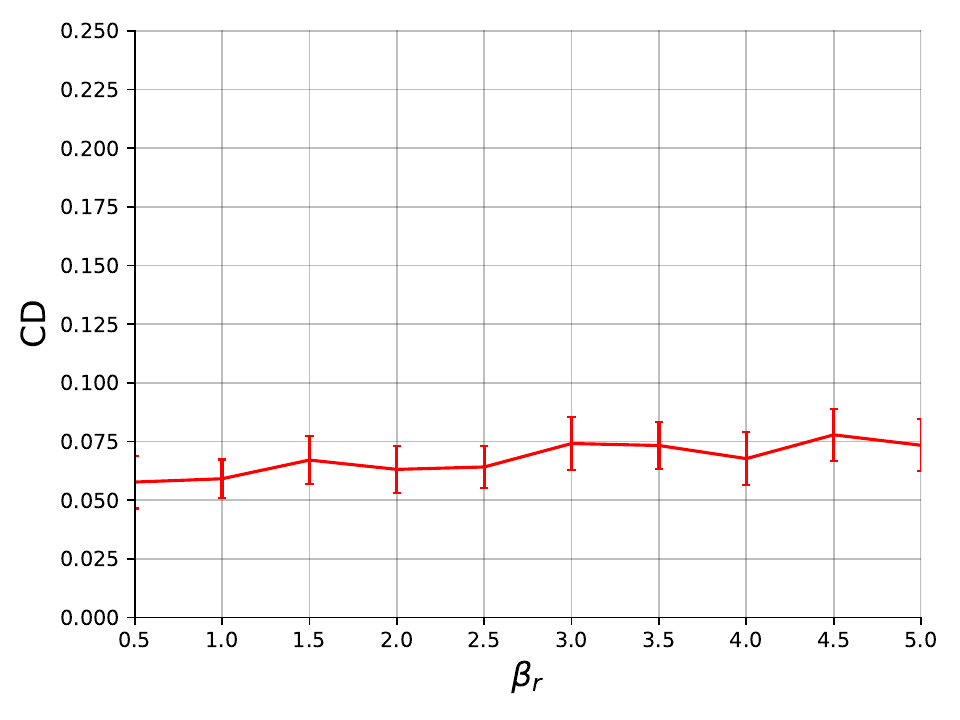}
        \caption{Spatial infection radius (\(\beta_r\)) - \(CD\).}
        \label{fig:sens_beta_cd}
    \end{subfigure}
    
    \caption{A sensitivity analysis of the average reproduction number \((E[R_t])\), max infection (MI), and cows death (CD) portion. The results are shown as mean \(\pm\) standard error of \(n = 1000\) simulation realizations. The y-axes are identical for all panes of a given row. In general, exploration level (middle column) shows a more pronounced effect on all three indices compared to the number of pathogens (left column) and spatial infectious radius (right column). }
    \label{fig:sensitivity}
\end{figure}

Moreover, we investigated the sensitivity of the epidemic spread dynamics with respect to the main model parameters. To this end, Fig. \ref{fig:sensitivity} summarizes the main sensitivity results of the proposed model where the x-axis indicates a focal parameter of interest and the y-axis indicates the epidemic spread metrics' values. Results are shown as a mean \(\pm\) standard error of \(n=1000\) simulation realizations. Notably, each simulation realization uniformly samples the model's parameters from the ranges presented in Table \ref{table:parameters}, ensuring each realization is unique. As can be seen from Figs. \ref{fig:sens_k_e_r_0}, \ref{fig:sens_k_mi}, and \ref{fig:sens_k_cd}, all metrics generally increase with the number of pathogens \((k)\). To be exact, MI only increases in low values of \(k\) and then reaches a plateau. Focusing on the ARN (\(E[R_0]\)), the standard error of the results also increases indicating that the system becomes more chaotic and less predictable, with realized ARN more sensitive to the specific parameters of each realization. The MI metric seems to reach an asymptote or even a peak around \(0.44\) at \(k=5\) and then stops increasing and even slightly decreases. This behavior might reflect that the infected sub-population dies faster than it has an opportunity to further spread the pandemic, on average, emphasizing why consideration of multiple pathogens in concert can overturn the result of simpler SIR models for single pathogens. Lastly, the CD presents relatively stable behavior of monotonic increase with respect to \(k\), as indicated by its error bars. 

In a similar manner, when the exploration to exploration walk ratio (\(\omega_1/\omega_2\)) increases, as shown in Figs. \ref{fig:sens_omegas_e_r_0}, \ref{fig:sens_omegas_mi}, and \ref{fig:sens_omegas_cd}, all three epidemic spread metrics also monotonically increase. The ARN shows similar behavior as before while the MI and CD metrics indicate a polynomial (non-linear) increase. Moreover, the CD was shown to become less stable and higher as the exploration-to-exploration walk ratio increased. Lastly, the spatial infection radius (\(\beta_r\)) increases and the ARN also slightly increases in a linear manner, keeping relatively the same level of stochastic behavior. On the other hand, the MI and CD metrics seem to be non-linearly affected by the spatial infection radius as they slightly increase and decrease for different values of \(\beta_r\). 

\begin{figure}[!ht]
    \centering
    \begin{subfigure}{.32\textwidth}
        \includegraphics[width=0.99\textwidth]{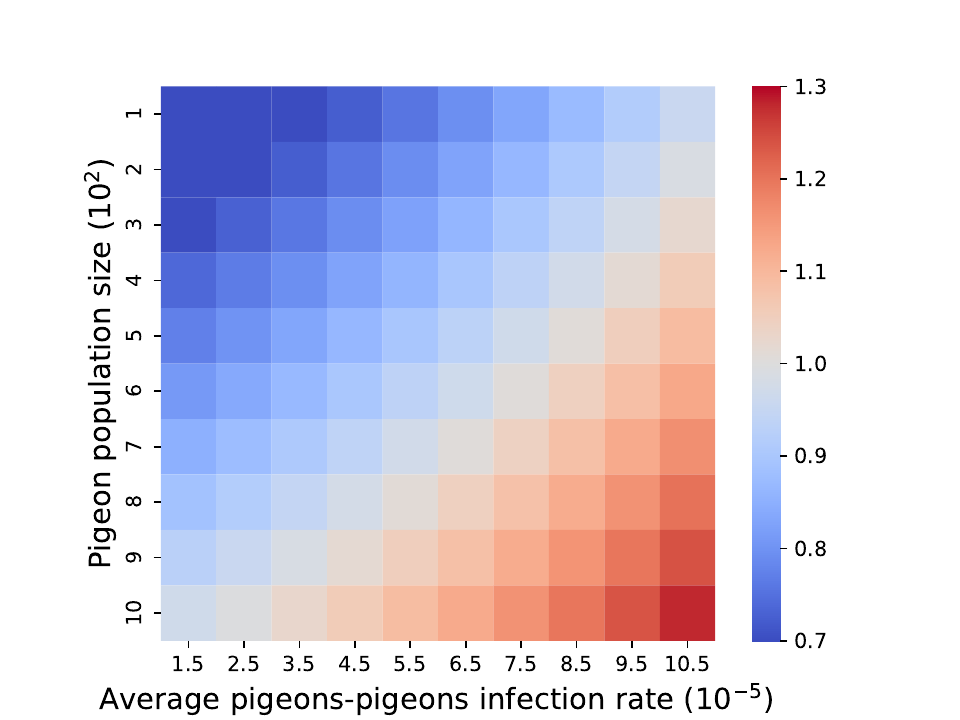}
        \caption{ARN (\(E[R_0]\)).}
        \label{fig:r_zero}
    \end{subfigure}
    \begin{subfigure}{.32\textwidth}
        \includegraphics[width=0.99\textwidth]{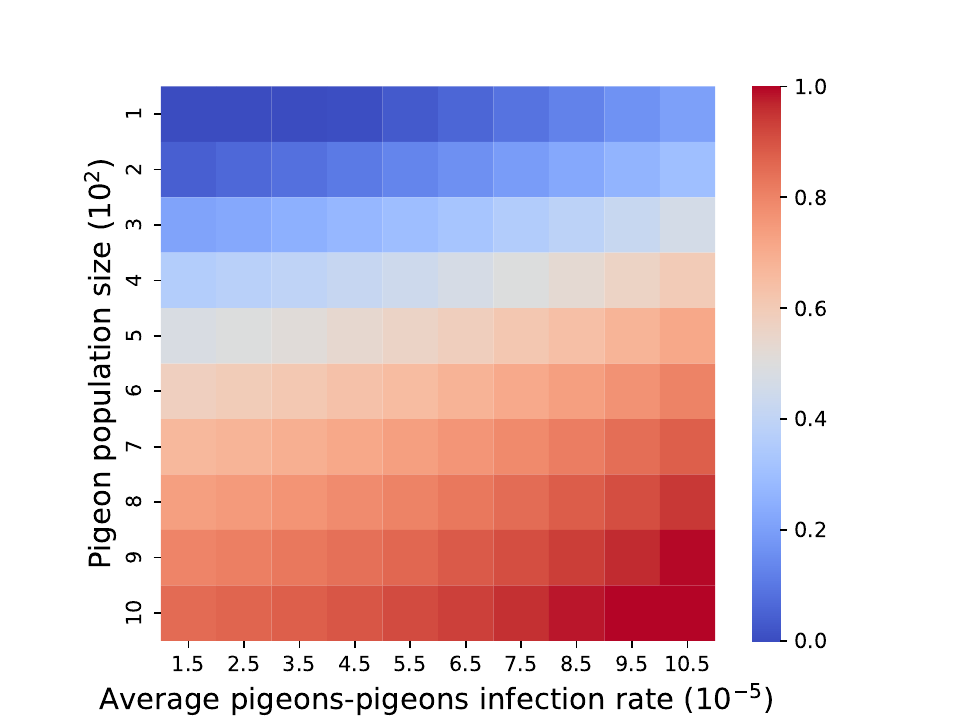}
        \caption{Max infected (MI).}
        \label{fig:mi}
    \end{subfigure}
    \begin{subfigure}{.32\textwidth}
        \includegraphics[width=0.99\textwidth]{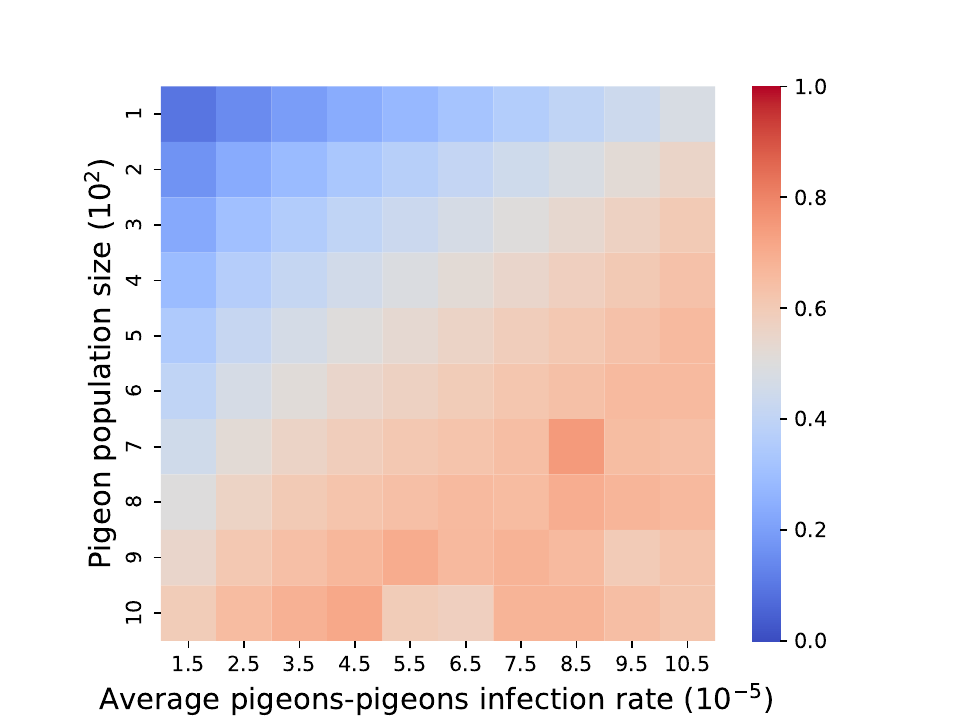}
        \caption{Cow death portion (CD).}
        \label{fig:cd}
    \end{subfigure}
    
    \caption{A sensitivity analysis of different epidemic spread metrics as a function of the initial pigeon population size (\(|P(0)|\)) and average within species infection rate (\(\beta\)). The results are shown as the mean value of \(n = 1000\) simulation realizations. The results show an interactive and non-monotonic effect of the two predictors (population site and infection rate) on disease outbreak indices.}
    \label{fig:heatmaps}
\end{figure}

In addition, in order to investigate the influence of pigeons as epidemic carriers, we computed the epidemic spread of the proposed scenario as a function of the initial pigeon population size (\(|P(0)|\)) and their average within-species infection rate (\(\beta\)). The results of this analysis are summarized in Fig. \ref{fig:heatmaps} such that each heatmap indicates a different epidemic spread metric. First, Fig. \ref{fig:r_zero} reveals that the ARN (\(E[R_0]\)) ranged between 0.7 and 1.3 indicating the interactive effect of the two aspects, so that for small pigeon populations even if these very sick or large pigeon populations with mildly infective pathogens would not cause an outbreak while the combination would. Second, Fig. \ref{fig:mi} shows that the MI metric where there is a second-order polynomial relationship between (\(|P(0)|, \beta\)) and the MI metric: \(MI = 0.172 + 0.032|P(0)| + 0.019\beta + 0.001\beta^2 - 0.001|P(0)|^2 - 0.001|P(0)|\beta \). This fitting is obtained using the SciMed symbolic regression tool \cite{scimed} with a coefficient of determination of \(R^2 = 0.72\). Finally. Fig. \ref{fig:cd} presents a linear increase toward \(CD = 0.58\) followed by a Plato with noisy results. This outcome indicates that a more aggressive pathogen with larger populations causes more spread in the short term but decays faster, leaving less overall mortality rate \cite{teddy_economy}.

\section{Discussion}
\label{sec:discussion}
In this study, we address the spatio-temporal dynamics of multi-pathogen epidemic spread and apply a general epidemiological model to a specific study system where pigeons serve as vectors of pathogens among dairy farms, transmitting avian (pigeons) and mammalian (cows) diseases. The model follows recent extensions of the well-established SIR modeling approach \cite{first_sir} into a multi-species with parallel multi-pathogen dynamics, within a Susceptible-Exposed-Infected-Recovered-Dead (SEIRD) framework. We implemented the model as an agent-based simulation approach using the data collected by \cite{orr_base_paper} as well as relevant epidemiological parameters from the biological literature (e.g. pathogen spread dynamics). This parametrization step allows us to test a realistic setup for the proposed model. The simulations show that considering the unique movement patterns of pigeons and their potential role as vectors can generate different outbreak dynamics than predicted otherwise. Further, we also show that variation among individual pigeons is not only affecting outbreak indices in accordance with previously published results for the importance of super-spreaders for disease dynamics \cite{dis_or_1,dis_or_2,dis_or_3} - it also affects the stability of the system and the predictability of the results. These two effects, together with evident interactions resulting from co-infections, demonstrate the utility of our model for predicting more realistic outbreak dynamics. This is supported also by previous studies establishing extended SIR-based models' ability to accurately capture similar epidemiological cases with only partial data \cite{multi_populations_4,multi_species_sir_model,sir_example_2,sir_example_9}. Below we first discuss the impact of pigeon movement on the predicted disease dynamic, then we turn to discussing the effects of other parameters and the added value of considering multiple pathogens in the same model. We conclude by pointing out some possible limitations of our model, directions for future studies, and the broader impacts of the SIR modeling approach.

\subsection{Vector movement and its variation affect outbreak dynamics and predictability}

The proposed model shows that pigeons operate as an effective epidemic spread vector, as illustrated by Fig. \ref{fig:cowsheds}. When the pigeons' population size is small, even if an epidemic outbreaks in one cowshed, it is not likely to be transformed into other cowsheds. While this outcome may be considered (almost) trivial, we show that the addition of a growing number of cowsheds does not increase linearly as one may obtain from a classical SEIRD model \cite{sir_review} due to the exploitation behavior in the pigeons' spatial dynamics (tendency to revisit a familiar site) which keeps some level of separation. This separation, on average, causes a less aggressive epidemic spread compared to a more naive model. This paints an interesting outcome- namely, pigeons during non-outbreak do not cause much epidemiological harm if the cowshed is far enough from other cowsheds. In a complementary manner, the epidemic caused by pigeons, especially in multi-cowshed scenarios, is self-controlled as the reduction in the pigeons' movement reduced the pathogen spread over time, as shown in Fig. \ref{fig:sensitivity}. That said, this result should be taken with caution as the outcome achieved for a simplified case of pathogen-movement relationship which is not biologically validated yet. Namely, it is known that pathogens indeed often reduce movement to a varying extent, but in some scenarios may in fact enhance movement through host manipulation \cite{con_new_1}. 

We also find that as the exploration-to-exploitation walk ratio increases, the maximum portion of infected animals increases polynomially fast (Fig. \ref{fig:sensitivity}). This phenomenon happens as the random walk takes a larger portion of the pigeons' movement making them closer to a well-mixture scenario for at least the during periods of the day when they either nest together or visit the cowshed \cite{alexi_games,dens_1}. From a modeling point of view, when (\(\omega_1/\omega_2\)) is relatively small, one can approximate the spatial dynamics using a graph-based model rather than a continuous one as the pigeons would visit a finite set of locations over time while the transformations between them do not have much effect on the infection dynamics \cite{reviewer_1,graph_based,teddy_sir_review}. Moreover, the spatial infection radius seems to have only a minor influence on the overall epidemic spread, as revealed in Figs. \ref{fig:sens_beta_e_r_0}, \ref{fig:sens_beta_mi}, and \ref{fig:sens_beta_cd}. This outcome can be associated with the fact that with high enough density such as the one present in the cowsheds, the infection radius does not have much effect as shown by previous models \cite{dens_2,dens_3,dens_4}. in other words, this small-scale variation in spatial scale is only minor compared to the local density of cows and movement of the pigeons (vector) \cite{intro_0}. 

Perhaps our most striking results are the consequences of individual variation among pigeons in their movement. Such behavioral variation is receiving growing attention in the ecological literature, with accumulating evidence for the generality of the pattern across animals \cite{con_new_2}, and for its potential impact on various system-level outcomes, including contact networks and disease dynamics \cite{pop_orr_1,con_new_3}. In our model, as presented by Fig. \ref{fig:last_fig}, increasing variation among individuals in their exploration-exploitation tendencies resulted in higher ARN and MI and eventually more cow mortality. This result concurs with existing literature (both models and empirical case studies) highlighting the role of super-spreaders in facilitating disease transmission \cite{dis_or_1,dis_or_2,con_new_4}. Second, it resulted in increasing variation among realizations, implying lower stability of the results and enhanced unpredictability of the dynamics. With increasing variation, an outlier individual is more likely to connect otherwise isolated sites or contribute to outbreak dynamics in an extraordinary manner. Thus, the strength of mean-field estimates becomes less certain, with higher sensitivity to stochastic conditions and various parameters (note that we randomly selected parameter values for each iteration). Despite this intuitive interpretation, we are unaware of empirical examples demonstrating this effect, highlighting the novelty of this prediction and the overall value of our modeling approach.   

\subsection{Other factors affecting outbreak dynamics and the benefit of multi-pathogen modeling}

In addition to \textit{in silico} experiments on pigeon movement, we explore the influence of different biological properties of the system on the epidemic spread. Specifically, Fig. \ref{fig:sensitivity} shows the sensitivity of the average reproduction number (ARN), max infected (MI), and cow death portion (CD) metrics as a function of the number of pathogens (\(K\)), and spatial infection radius (\(\beta_r\)). Interestingly, as the number of pathogens increases up to five, the number of individuals infected at the same time increases to around half of the population, as one can see from Figs. \ref{fig:sens_k_e_r_0}, \ref{fig:sens_k_mi}, and \ref{fig:sens_k_cd}. Nonetheless, after this point, as the number of pathogens increases the MI metric remains constant. This outcome is associated with the following dynamics: one pathogen has over the other as too many pathogens cause a higher mortality rate, they actually reduce the overall infection rate as individual die quicker and does not have an opportunity, on average, to infect other individuals in the population \cite{teddy_multi_strain,multi_strain_1,multi_strain_2}. 

From the proposed analysis, it seems that when the pigeons' population size is controlled, they do not cause epidemic outbreaks in their own population or the cow population in cowsheds. However, if the pigeon population size increases or if a set of multiple, highly contentious pathogens is introduced, an epidemic outbreak is only a question of time. Importantly, if no epidemic intervention policies are quickly and efficiently used, then once just several cows have been infected, most of the cow population within that site would be infected within a small period, due to the small spatial location they share  \cite{graph_2,graph_4}. In contrast, in multi-cowshed scenarios, it seems that even without an intervention, the system will mitigate the outbreak and the epidemic spread over a (relatively) long period of time. This conclusion highlights the complexity of the biological dynamics in such agricultural settings as these have self-stabilizing properties on the one end but are extremely sensitive to outside influence. Therefore, lacking more epidemiological data, it is challenging to draw any definitive conclusions.  

When focusing on the influence of the pigeons as an external species in the context of a cowshed, the epidemic spread can be reduced to two main factors - the pigeons' population size and average within-species (instra-specific) infection rate. Following this simplification, the results show (Fig. \ref{fig:heatmaps}) the influence of these two parameters on the ARN, MI, and CD. Fig. \ref{fig:cd} shows that even a relatively small pigeon population with a relatively non-aggressive infection rate can cause cows' deaths if not treated but this would be minor, on average. On the other hand, only a larger pigeons' population and a more aggressive infection rate result in a global outbreak as indicated by Fig. \ref{fig:r_zero}. Taken jointly, the results show that pigeons can cause sporadic infection and death in cows even in small population sizes but it requires both larger pigeons' population size and aggressive infection rate pathogens, and that the pathogen(s) will not have a strong negative effect on their movement (Fig. \ref{fig:sick_movement}). That said, it is important to note that even for extreme cases such as 1000 pigeons with highly infected pathogens, the average reproduction number reaches only 1.3 which is comparably small to other scenarios such as the Zika and COVID-19 epidemics \cite{review_zika,covid_with_ventilation,review_covid}. 

\subsection{Future direction and concluding remarks}
While our model presents a comprehensive approach to studying the pigeons in the cowshed epidemic spread settings, there are several limitations to the proposed model and analysis. First, the accuracy of the model heavily relies on the quality and representativeness of the data used for calibration and validation. Namely, we only partially establish the model's effectiveness. Whereas we have solid spatial data for parameterization, we are still lacking the epidemiological data to fully validate it. Gathering reliable data on animal interactions and disease prevalence can be challenging (particularly in field settings), and future work may focus on collecting such data and re-evaluating the performance of the proposed model. Second, our model simplifies several biological attributes of the system. For instance, we assume a continuous birth of pigeons over time, ignoring the breeding biology of simultaneous clutches that can alter the dynamics. Similarly, we assume random encounters within the population, while social structures and the dependency between individual movement and the subset of individuals it interacts with can affect effective beta and ARN \cite{or_sir_example,con_new_2,intro_1}. Third, the proposed model does not consider the potential genetic variability of pathogens circulating within the pigeon and cow populations. Previous studies show that over time, taking into account the genomics mutation process of pathogens plays a critical role in understanding epidemic spread \cite{pandemic_mutate,teddy_gaddi,polio_mutation_rate,most_mutation_die}. As such, incorporating better parameters, more biological realism, and genetic data into the proposed model could provide a more nuanced understanding of pathogen interactions.

Finally, despite these above-mentioned limitations, our model highlights the premise of combining extended SIR models (here spatio-temporal multi-pathogen SEIRD) and simulations for addressing concurrent challenges related to disease transmission in agricultural and urban settings. Such models, when coupled with empirical data can refine predictions regarding systems and pathogens of interest. Simulation and model analysis can facilitate the evaluation and prioritization of effective interventions, before their practical (pricey) implementation (e.g. should one reduce pigeon population size through culling or mobility through fencing?). These models and their predictions can directly serve the \say{One Health} approach \cite{con_new_5,con_new_6,con_new_7}, highlighting the urgent need to link wildlife behavior, agricultural practices, and human health. In a world suffering from an over-accelerating rate of zoonotic diseases such models are essential approaches for capturing the complex dynamics of these interdependent systems.

\section*{Declarations}
\subsection*{Funding}
OS acknowledges financial support by Grant 891-0232-21 from the Israel Dairy Board Research, grant ISF396/20 from the Israeli Science Foundation, by the Data Science Center at Tel Aviv University, and by the Koret-UC Berkeley-Tel Aviv University Initiative in Computational Biology and Bioinformatics.

\subsection*{Conflicts of interest/Competing interests}
None.

\subsection*{Code and Data Availability}
The code and data that have been used in this study are available upon reasonable request from the authors.

\subsection*{Acknowledgment}
We gratefully thank Miranda Crafton for her guidance, Shay Cahani for his biological consulting, and Avishai Lublin for his valuable feedback and discussions.  

\subsection*{Author Contribution}
Teddy Lazebnik: Conceptualization, Software, Formal Analysis, Investigation, Methodology, Visualization, Project administration, Writing - Original Draft, Writing - Review \& Editing. \\ 
Orr Spiegel: Conceptualization, Resources, Data curation, Validation, Investigation, Writing - Review \& Editing. \\
 
\bibliography{biblio}
\bibliographystyle{unsrt}

\end{document}